\documentstyle{laa}
\input epsfig
\begin{document}
%
\thesaurus{13. 
    (02.07.1,02.09.1,05.03.1,11.05.2)}

\title{On the stability of motion of $N$-body systems: a geometric approach}
\author{A.A. El-Zant}
\author{A.A. El-Zant$^{1,2}$}
\institute{Astronomy Centre, University of Sussex,
Brighton BN1 9QH, UK\and Physics Department, Technion --- Israel Institute of Technology, Haifa 32000, Israel}
\date{Received .........; accepted.........}
\maketitle

\begin{abstract}

Much of standard galaxy dynamics
rests on the implicit assumption that the corresponding 
$N$-body problem is (near) integrable. This notion although leading to great
simplification  is by no means a fact. In particular, this assumption is unlikely to
be satisfied for systems which display chaotic behaviour which
manifests itself on short time-scales and  for most initial conditions. 
It is therefore important to develop and test methods
that can characterize  this kind of behaviour in realistic situations.
 We examine here a method, pioneered by Krylov (1950) and first introduced to 
gravitational systems by Gurzadyan \& Savvidy (1984,1986). It
involves a metric on the configuration manifold 
which is then used to find local 
quantification of the divergence of trajectories and  therefore appears to be
suitable for short time characterization of chaotic behaviour.
 We   present  results of  high precision $N$-body simulations of the dynamics
of systems of 231 point particles over a few dynamical times. The 
 Ricci (or mean) curvature  is calculated along the trajectories.
Once fluctuations due to  close encounters are removed this quantity is found to be
almost always negative and therefore all systems studied display local
instability to random perturbations along their trajectories.
However
it is found that when significant softening is present the Ricci curvature is no longer
negative. This suggests that smoothing significantly changes the structure of the $6N$ phase
space of gravitational systems and casts doubts on the continuity of the transition
from the large-$N$ limit to the continuum limit.
 From the value of the negative curvature, evolution  time-scales of systems displaying clear 
instabilities (for example collective instabilities or violent
relaxation) are derived. 
We compare the predictions obtained from these calculations with the time-scales
of the observed spatial evolution  of the different systems and deduce that this 
is fairly well described.
In all cases the
results based on calculations of the scalar curvature qualitatively agree.
These results suggest that  future
applications of these methods to realistic 
systems may be useful in characterizing their stability properties. 
One has to be careful however in relating the time-scales 
obtained to the time-scales of energy relaxation since different dynamical quantities
 may relax at different rates.
\end{abstract}

\section{Introduction and motivation}
\subsection{On the assumptions of standard galaxy dynamics}

  The  classical assumption often made in galaxy dynamics (Binney
\& Tremaine 1987 (BT) chapter 4) is that 
present day galaxies can be treated as collisionless fluids in steady states
described by one particle distribution functions obeying the time independent 
collisionless Boltzmann equation (CBE). In addition, it is often supposed that 
these distribution functions are completely characterized by the isolating
integrals of motion (of single stars).

implies that almost all orbits are regular and therefore stars conserve as many 
integrals of motion (in involution)
 as the number of spatial dimensions they move in. Thus,
for a three dimensional system of $N$ stars there are $3N$ conserved quantities
and the $N$-body problem is solvable by quadratures---or integrable (see e.g., 
Whittaker 1937 or Goldstein 1980 for discussions using classical analysis: more modern 
treatments can be found in Arnold 1989 (ARN) or Abraham \& Marsden 1978). 

 Thus in this picture galaxies are modeled as
 classical fluids with time independent 
 densities which give rise to potentials similar to those
for which the Hamilton-Jacobi equation is separable.
Obviously these assumptions cannot be strictly satisfied for real 
galaxies.
It is argued however, that due to the large two body relaxation 
time (e.g., Saslaw 1985)
\begin{equation}
   \tau_{b}=\frac{v^{3}}{32\pi G^{2}m^{2}n\ln(N/2)} \label{eq:twobod}
\end{equation}
(where $v$ is the characteristic speed of a star and $n$ is the number density of
``field stars'' of mass $m$) that these systems can be treated as effectively collisionless 
over many Hubble times. 
However all that Eq. (1) is saying is that the direct impulse from encounters
between pairs of stars is small relative to the velocity of a star in
the smoothed potential. 
This is due to the long range nature of gravitational interactions
which ensures that the whole system contributes to the mean force on a star
at all times.
In the approximation leading to (1) however
the perturbation to the trajectory of a test star due to a certain field
star is only calculated at their closest approach --- that is only once during
a crossing time.
Nevertheless, due to the complicated nature of 
the solutions of the Newtonian equations for $N>2$, discreteness can have a
major indirect effect; the nonlinearity of the problem prevents the
adding up of the motions due to 
individual binary interactions to each other and to the motion in the 
mean field since although the forces add up linearly the solutions of the Newtonian
equations do not.
This suggests that  
the $N$-body problem is perhaps best studied in its entirety.
In this regard  it may be perhaps useful to recall that
 the collisionless steady-state approximation 
is not a trivial simplification of the dynamics: It reduces the $N$-body
problem to $N$ one-particle problems in a given potential, thus  
reducing the
number of coupled first order differential equations to be solved 
from $6N$ equations to $6$ equations.

For general Hamiltonian dynamical systems, regular motions form invariant
(under time propagation of the solution) tori in the $6N$ phase-space. In 
integrable systems these occupy the whole of that space. Under perturbations,
however, some tori are destroyed leaving behind volumes of phase space
 where irregular (or chaotic) 
motion can occur. The extent of the chaotic region will depend on the 
strength of the perturbation. Bounds given by the KAM theorem predict that a 
positive measure of tori will survive provided the perturbation to the potential
$\mu<\mu_{c}$, where $\mu_{c}$
is a critical amplitude (e.g., Mackay \& Meiss 1987). In general, the value of $\mu_{c}$ decreases
as $\exp(-N \log N )$ and has been shown to be irrelevant for many higher
dimensional physical systems (Pettini 1993 (P93) and the references therein)
while the perturbation due to discreteness noise decreases only as
$1/\sqrt{N}$ if it is random (e.g., Saslaw 1985). 
The diffusion time away from the remaining tori is also a decreasing
function of $N$; $\tau_{D}\sim \exp(1/\mu)^{1/N}$ (Nekhoroshev 1977;
Perry \& Wiggins 1994). These results 
suggest that a system is unlikely to become more regular with increasing $N$,
 in direct contradiction with the 
 results of two body relaxation estimates but as expected from
considering the complicated nature of solutions of generic $N$-body systems. In fact,
for the particular case of a spherical gravitational system, it was shown by
Gurzadyan \& Savvidy (1986) (GS) that as $N$ increases 
this system tends towards an Anosov (1967) C-system
with maximal instability in phase-space (if one neglects escapes and direct collisions).
Making the same kind of assumptions as those used to obtain Eq. (\ref{eq:twobod}) (by considering an infinite and homogeneous medium), GS obtain the following
relation for the $N$-body relaxation time arising from the instability
\begin{equation}
\tau_{gs}=(15/4)^{2/3}\frac{1}{2 \pi \sqrt{2}} \frac{v}{G m n^{2/3}}.
\label{for 2}
\end{equation}
which is considerably shorter than the binary relaxation time.

 C-systems are very irregular  --- to the point that their evolution can be described
as a Markow process (e.g., Pesin 1989).
The collisionless approximation however predicts that
spherical systems end up in completely integrable steady states --- the most
regular and predictable systems that can exist.
The contradiction can possibly arise because the  analysis leading to (1)
focuses on the fact that the force function becomes smoother as $N$
increases. It then assumes directly that this implies the solution becoming 
more regular  not 
taking into account that exponentially smaller perturbations
are sufficient with increasing $N$ to make an $N$-body system unstable.
 Therefore, while some quantities that do not depend on the exact details
of the dynamics may be slowly evolving and not sensitively dependent on
the discreteness noise, the solutions themselves are heavily sensitive
to noise.
 For example, the change in the energies of
stars that arises from discreteness noise is likely to be much slower
than the changes in their trajectories. A rough example  is the
fact that numerically integrated orbits in fixed (and smooth) potentials
can have the energy (and all the Poincar\'e invariants) 
conserved along their trajectories to up to ten
digits while the trajectories themselves are completely inaccurate 
perhaps not resembling (even qualitatively) the real ones (El-Zant 1996b). 
The numerical
errors here represent the noise (if they are taken to be random,  a 
proposition still under discussion: McCauley 1993).  This is because
here energy is a scalar function of the phase space variables and is related
to the force by a path independent integral. That path on the other 
hand is a 6 component vector in phase space. One can thus easilly envisage
perturbations that change the trajectory of a particle without changing its energy.    
 A long time-scale of energy relaxation does not therefore imply that the actual detailed 
dynamics are collisionless in the sense of being stable solutions of the CBE. 
One would expect that some macroscopic quantities 
could be affected, for example velocity dispersion or the actual shape
of the gravitational system.

In fact, simulations of single particle motions
in fixed potentials show that the time-averaged phase-space density distribution
of individual chaotic trajectories, as well as their statistical
properties, change significantly over a time-scale $\ll\tau_{b}$ when the
potential is given some graininess (Pfenniger 1986; Udry \& Pfenniger 1988), or when  the trajectories are given small random kicks without changing their total energies significantly
(Kandrup 1994; Merritt \& Valluri 1996; El-Zant 1996b). 
The full $N$-body problem  would be expected to be much 
more irregular and prone to 
evolution on small (compared to $\tau_{b}$) time-scales. 
 Such effects are actually observed in $N$-body simulations
of up to $10^{6}$ heavily softened particles but are assumed to be due to
the relatively small number of particles present, even though the two body relaxation time in these simulations is still usually much larger than  the time-scales
considered. 
(a
review of such occurrences is given by Hernquist \& Ostriker 1992: see also van Albada 1986; Sellwood 1987: Zhang 1996 presents a detailed study of an example of a case where
discreteness noise interacts with global irregularities in the density thus triggering evolution). 
In addition, processes involving 
 non-stellar objects such as interactions with
giant molecular clouds in galactic disks or
black holes in halos, small dissipative perturbations (e.g., Pfenniger \& Norman 1990)
 etc.  can cause even more serious trouble for the 
collisionless approximation.

 Even if the collisionless approximation does hold, this
does not guarantee  that  a given density distribution would not evolve
over a Hubble time. 
This is because individual trajectories, even in a smooth steady-state
potential, can have time dependent density distributions over such a time-scale. 
This would be the case in general non-spherical 
potentials as noted by Binney (1982). Hasan  et al. (1993) describe such
behaviour for barred spirals while  Merritt \& Fridman (1996) show that
it may also be important for ellipticals. El-Zant (1996b) examines the case 
of trajectories started near the symmetry plane 
of disks embedded in triaxial halos. 
 Finally, it seems that perhaps there is also some observational evidence
for evolutionary phenomena occuring in galaxies
 (Wielen 1977; Pfenniger et al. 1994; Courteau et al. 1996; Pucacco 1992). 

The above considerations suggest (although in no way prove)  
that even though gravitational systems obey the CBE in  the infinite $N$ limit 
(e.g., Braun \& Hepp 1977)  the 
existence and stability (against discreteness noise)
of steady state solutions of that equation are in question when a significant 
amount of chaos is present. This means that while in some cases the classical theory may still 
 hold over a few Gyr
(considering the primitive evolutionary state of cold disks in some spiral
galaxies one has to admit that this could often be the case) 
it does not always obviously do so.

\subsection{How chaos drives evolution}

A distribution of stars giving rise to an 
integrable potential can either oscillate 
coherently or reach a macroscopic steady state through phase
mixing. This is a trivial form of relaxation which is simply due to stars moving at
different angular frequencies on their respective KAM tori in the 
6-dimensional phase-space --- the motion in  the $6N$-dimensional phase-space
being a $3N$ torus characterized by the $3N$ integrals. Phase mixing conserves the
action variables characterising this torus and which are crucial in determining the physical state of a system.
 Most galaxies
are not believed to be undergoing significant
large scale oscillations, we therefore conclude
that if they are integrable dynamical systems they must be in a steady state.
Moreover, if they are sufficiently far from any  chaotic system
they must be stable to small perturbations (KAM theory). This is what is assumed
in the classical theory.

A necessary condition for evolution therefore is
the non-integrability of the system. This condition is satisfied for $N$-body gravitational
systems since there are no global integrals of motion other than the 
classical  ones (Poincar\'e 1889).
However, not all non-integrable
systems exhibit interesting behaviour (different from classical theory) in
the time-scales of interest. We need the property of {\em phase-space}
mixing --- that is the spread of 
localized volume elements (corresponding to sets of 
initial conditions) to cover large areas of the $6N$ phase-space (while still
conserving their original Lebesgue measure: e.g., Sagdeev et al. 1988). 
Phase-space mixing leads to diffusion in the action variables and may therefore
lead to evolution in a system's physical parameters. 
For this process to be efficient it is necessary that the system be sufficiently chaotic so that the diffusion 
occurs over short enough time-scales and covers a large range of initial 
states, only then  can we say that the macroscopic state corresponding to a
given set of micro-states can evolve over a Hubble time.

It is therefore important to 
examine different methods for detecting these processes 
and the time-scales associated with them in practical situations where
the predictions can be tested. 
The rest of this study is a  step in this direction.
In the next section we describe why certain subtleties related
to $N$-body gravitational systems require {\em local} methods to be used
in the quantification of chaotic behaviour.
One such approach based on the geometry of the configuration manifold 
is then described. In section 3 we describe some of the possible applications of this approach while in section 4 results of some
numerical experiments testing the method are reported. 
\section{Characterization of chaos and the Ricci criterion}
\subsection{Difficulties with commonly used methods}
Central to the idea of phase-space mixing is the 
concept of dynamical entropy used to quantify it. The most important of such quantities
is the so called Kolmogorov-Sinai entropy. 
The easiest 
way of calculating the KS entropy is by evaluating the Liapunov exponents,  
which for a dynamical system defined by the vector equations
\begin{equation}
{\bf \dot{X}=F(X)},       \label{eq:linf1}
\end{equation}
with solution   
${\bf \bar{X}=\bar{X}}(t,t_{o},{\bf \bar{X}_{o}})$ 
are given by
\begin{equation}
\sigma(\mbox{\boldmath$\xi_{0},X_{0}$})= \lim_{t \rightarrow \infty} \frac{1}{t}
\log
\frac{\parallel
\mbox{\boldmath$\xi$}(t)\parallel}{\parallel\mbox{\boldmath$\xi$}
(0)\parallel}       \label{eq:linf4},
\end{equation}
where {\boldmath$\xi$} is a tangent space vector arising from the solution of
the variational equations
\begin{equation}
\dot{\mbox{\boldmath$\xi$}}={\bf D_{x}F} \left(
{\bf \bar{X}}  \left(t,t_{o},{\bf X_{o}} \right) \right)
    \mbox{\boldmath$\xi$}.  
\label{eq:linf3}
\end{equation}
The KS entropy is then usually given by (e.g., LL)
\begin{equation}
KS=\int \sum_{i} {\mbox {\boldmath $\sigma(\xi^{i},X_o)$}} d{\bf X_o},
\label{EQ}
\end{equation}
where the sum is taken over all positive exponents and the integral
is over all possible initial conditions.
For systems with simple enough phase-space and when the infinite time characteristics are required,
the KS entropy calculated with the help of this formula suffices to describe  the
ergodic properties of a system. Positive KS entropy over
a compact phase-space, or some region of it that has this  property,
is a sufficient condition for the presence of what is usually called chaos and
the accompanying erratic behaviour which leads to the approach towards
statistical equilibrium  even in low dimensional systems (e.g., McCauley 1993).
The evolution time-scale is usually related to the inverse of the KS entropy.

Open systems interacting via un-softened Newtonian potentials do not have
compact phase-spaces however, and therefore the situation is more complicated.
Here, no final state exists and one 
has to distinguish between the various 
stages of evolution i) Violent relaxation, 
ii) Collective (plasma type) instabilities iii) Evolution 
towards an isotropic rotator and
iv) Kinetic evolution towards equipartition of energy, core 
collapse etc. Here, the distinctions are 
rather arbitrary and are used only for clarity --- these processes may interact and influence each other. 
Stage three has been given very little attention in the 
literature because of the lack of a mechanism from 
traditional physics leading to such evolution 
(on a time-scale smaller than given by (\ref{eq:twobod})). 
Nevertheless, the chaotic nature of the $N$-body problem may 
provide a clue as to how this might happen.

All the above processes should be characterized by  phase-space instability leading to mixing, but evidently cannot be described by any quantities defined only for
infinite times. One way out of this problem is to integrate either the 
linearized equations (\ref{eq:linf3}) (e.g., Goodman  et al. 1993) or the full 
non-linear equations (\ref{eq:linf1}) for slightly different initial conditions
(e.g., Kandrup et al. 1994 and the references therein) for short times.
  One drawback of such an approach, however, is that one is comparing the 
divergence  in phase-space of different
temporal states and not trajectories. Integrable systems usually have linear
(in time) phase-space divergence between neighbouring states --- but {\em only on  average}. 
A simple illustration of the type of problems
involved is provided by examining the 
behaviour of a pendulum
\begin{equation}
\ddot{\theta}= - \sin \theta,
\end{equation}
with linearized equation
\begin{equation}
\delta \ddot{\theta}=- \cos \theta \delta \theta.
\end{equation}
When $\cos\theta$ is negative  this has a solution
\begin{equation}
\delta \theta \sim e^{t}.
\end{equation}
That is, during this interval, the solution of the linearized equation
predicts ``exponential instability''.  Obviously,
in this example, the trajectories do not diverge at all --- but the states did. 
In addition, methods that evaluate the whole set of Liapunov exponents 
of a dynamical system are fairly sophisticated (e.g., Eckmann \& Ruelle 1985)
and are not practical for higher dimensional systems. This in practice will
mean that one will have to evaluate only the largest exponent which is a measure
of the maximal instability in phase-space. Now, in a multidimensional separable
system  it is likely that at any given time
 there will be  some oscillations that are in the
``$\cos \theta < 0$ region'' --- that is displaying exponential divergence
in the linearized dynamics. An initial randomly oriented
vector in the linear tangent space will 
always reorient itself along the direction of maximum expansion under the stretching
effect of the flow (e.g., Wolf et al. 1985). It could therefore be possible to obtain average exponential
divergence  in the linearized dynamics even for separable systems.
This approach therefore may be unable to distinguish between genuine phase space mixing leading to evolution and trivial phase mixing of temporal states.  
It may also be useful to note here that there are few strict results concerning
the properties of the Liapunov exponents for higher dimensional systems 
(Eckmann \& Ruelle 1985; Pesin 1989).


\subsection{The study of motion on Lagrangian manifolds}

 There are a variety 
of ways of transforming Hamiltonian problems 
into the study of some metric space (see P93; Gurzadyan \& Kocharyan 1994 or the 
articles by Gurzadyan and Pettini in Gurzadyan \& Pfenniger 1994).
The oldest and most well known of these hinges on the observation, 
apparently first made by Hertz (1900) in the course of his 
remarkable reformulation of classical mechanics, that the 
Maupertuis principle  \begin{equation}
\delta \int_{\gamma} 2T dt=0, \label{eq:L1}
\end{equation}
(where $T$ is the kinetic energy along the motion on a trajectory $\gamma$) from which the equations of motion in their Lagrangian form arise is actually an expression for geodesics 
on a the configuration manifold $M$
where the motion is restricted as a result of conservation laws.
The fact that~(\ref{eq:L1}) defines geodesics is clear from the following relations: \begin{equation}
\delta_{e} \int T dt=\delta_{e} \int \sqrt{T} \sqrt{T} dt
=\delta_{e} \int \sqrt{T} dl=\delta_{e} \int \sqrt{E-V} dl,
\end{equation}
where $E=T+V$ is the total energy and $dl=\sqrt{a_{ij} dq^{ij}}$
with \begin{equation}
T=a_{ij} \frac{dq^{i}}{dt} \frac{dq^{j}}{dt} 
\end{equation}
for particles of unit mass.
Here the $a_{ij}$ are elements of the metric tensor and
$\delta_{e}$ refers to variations in the trajectory $\gamma$ holding the energy and the end points fixed. These 
are thus geodesics in the energy sub-manifold of the configuration space
and the $q^{i}$ are coordinates on it. If we now choose Cartesian coordinates in
the enveloping $3N$ space then \begin{equation}
dl^{2}=\sum  (dx^{\alpha})^{2}.
\end{equation}
The metric then is 
\begin{equation}
ds^{2}=\left(E-V \right) \sum_{3N} \left(dx^{\alpha} \right)^{2}. \label{eq:L2}
\end{equation}
From the Jacobi equation which describes the geodesic deviation on $M$ (e.g., Misner et al. 1973)
\begin{equation}
\nabla_{\bf u} \nabla_{\bf u} {\bf n + Riem(n,u)u=0},
\end{equation}
(where ${\bf n}$ is a separation vector analogous to $\mbox{\boldmath $\xi$}$ in~(\ref{eq:linf4}), ${\bf Riem}$
is the Riemann curvature operator, and 
$\nabla_{\bf u}$  the covariant derivative)
one can obtain  the following relation
for the norm of the {\em normal} component of this deviation
\begin{equation}
\frac{ d^{2} \parallel {\bf n} \parallel^{2} } { ds^{2} }
=-2k_{{\bf u,n}} {\bf \parallel n \parallel}^{2}+2{\bf \parallel
\nabla_{u} n \parallel }, \label{eq:L3}
\end{equation}        
where 
\begin{equation}
 k_{\bf u,n}=\frac{ \left[ \bf Riem(n,u) u \right] . \bf n }{\bf \parallel n \parallel}
\end{equation}
is the two dimensional curvature 
in a plane defined by ${\bf u \times n}$ and 
where we have used the fact that ${\bf n_{\bot}.u=0}$ . If $k_{\bf u,n}$
 is negative everywhere
for all planes as defined above 
(that is for all ${\bf n}$ normal to ${\bf u}$)
and if $-k=\min \mid k_{{\bf n,u}} \mid $ we have
 \begin{equation}
\parallel {\bf n} (s) \parallel \geq \frac{1}{2} \parallel
{\bf n} \parallel \exp \sqrt{-2k} s   \label{eq:L4}
\end{equation}
for ${\bf \dot{n} >0}$, and \begin{equation}
\parallel {\bf n} (s) \parallel \leq \frac{1}{2} \parallel
{\bf n} \parallel \exp - \sqrt{-2k} s, 
\end{equation}
for ${\bf \dot{n} <0}$. 
These relations describe the linearized dynamics in the ``dilating'' 
and ``contracting''
spaces characteristic of the class of 
Anosov (1967) C-systems to which, as was mentioned earlier, large spherical  $N$-body systems
belong. 
In this case,   $\sqrt{ -\sum k_{\bf u,n}}$ averaged over $M$ is the
KS entropy. In comparing two C-systems therefore  one can
{\em define the system with larger average value of $\sqrt{-\sum k_{\bf u,n}}$ to be
more unstable}. This system will have a larger exponentiation rate and so
initial conditions will tend to mix faster along geodesics. 
To determine how fast the
initial conditions mix in time, we note that $ds/dt=\sqrt{2} T$ so that
if $T$ does not vary too much during the evolution $s \sim \sqrt{2} T t$.
It is clear that a system with the above characteristics cannot conserve 
its action variables since there is always a deviation of trajectories
normal to the motion in phase-space (which is the cotangent bundle of $M$).

\subsection{Ricci curvature and the corresponding criterion}
GS have shown
that the condition for C-systems is not satisfied for general
gravitational ones. However, as Kandrup (1990a,1990b) has shown, the probability
of a two dimensional curvature being positive along a $N$-body system's trajectory
decreases exponentially with increasing $N$.
  Also, in the
$N$-body problem relation~(\ref{eq:L4}) implies that all orbits are unstable 
 at all times. However, one needs much less than 
what is described by~(\ref{eq:L4}) for observable effects of instability 
to be detected (just $10 \%$ of orbit becoming unstable may be enough).
On the other hand, just a few orbits being unstable in general would
not significantly change the physical  properties of a system.   
We therefore need some averaged form of~(\ref{eq:L3}) and a corresponding 
instability relation instead of~(\ref{eq:L4}) to characterize such behaviour.
Ideally such a relation should not require the evaluation of the Riemann tensor.

A natural way of proceeding is by using the Ricci (or mean) curvature of
the manifold $M$  
\begin{equation}
r_{\bf u}(s)=R_{ij} \frac { u^{i} u^{j} } {\bf \parallel u \parallel^2},  
\label{eq:L5}
\end{equation}
where $R_{ij}$ are elements of the Ricci tensor and $u^{i}=\frac{dx^{i}}{ds}$ are the
components of the geodesic velocity vector ${\bf u}$.
The Ricci curvature is related to the two dimensional curvatures by
 (Eisenhart 1926)
\begin{equation}
r_{\bf u}=\sum_{\mu=1}^{3N-1} k_{ {\bf n_{\mu},u} }(s).
\end{equation}
The value of $r_{\bf u}$ does not depend on the particular set of normal
directions ${\bf n}$ chosen so that $ r_{\bf u}/(3N-1)$ can be seen as the
average value of $k_{\bf u,n}$ over all possible directions normal to
${\bf u}$ on the configuration manifold $M$.   

In the case when all $k$'s are negative,
$\sqrt{-r_{u}}$ averaged over the whole manifold corresponds to the Kolmogorov entropy. In general, as was first noted by Gurzadyan \& Kocharyan (1987), it will provide  
an ``averaged'' measure of irregularity. In these terms 
 Equation~(\ref{eq:L3}) can be  written as 
\begin{equation} 
\frac{d^{2}Z^{2}}{ds^{2}}=  - 2 \left[\frac {r_{{\bf u}}(s)}{(3N-1)} \right] Z^{2} 
+2 \langle \parallel \nabla_{\bf u}{\bf n} \parallel \rangle,
\label{eq:L6}
\end{equation}   
where $Z$ is now to be interpreted  as  the norm of a vector that is
a member of a random field of vectors with uniform distribution in
directions normal to ${\bf u}$ and equal magnitude (El-Zant 1996a).

If $r_{\bf u}(s)$ is negative on a region of $M$
then one can obtain 
relations for $Z$ analogous to those obtained  for ${\bf \parallel 
{\bf n} \parallel^{2}}$  in~(\ref{eq:L4}) with $-k_{r}=\frac{\min[r_{\bf u}(s)]}{3N-1}$. One can then apply the
criterion of relative instability of C-systems to general Hamiltonian
dynamical systems in regions of their configuration 
manifolds where the Ricci curvatures are negative, 
which in this case will express the relative probability of any two systems being
unstable under random perturbations. We adopt here the following 
definition, convenient for numerical studies of $N$-body systems.

{\bf Definition}: \newline
\noindent
Let $R_{1}$ be some subset of the configuration manifold $M_{1}$ and $R_{2}$
be a subset of a manifold $M_{2}$ with $M_{2}$ not necessarily different from
$M_{1}$. Suppose also that the Ricci curvature is negative in both of
these regions. {\em We will say that $R_{1}$
corresponds to configurations of a dynamical system that are more
unstable than those represented by $R_{2}$ if
the average value of $\sqrt{-r_{\bf u}}$ is larger in $R_{1}$ than in $R_{2}$}.

{\bf Note that}: 
As in the case of C-systems we obtain time-scales from the relation
$\frac{ds}{dt}=\sqrt{2} T$.
If we are comparing systems with different kinetic energies,
the evolutionary times derived will have to be scaled accordingly. If
the kinetic energies of systems are changing in the region of the
dynamical systems of interest then time-scales derived from the Ricci
curvature alone
are not rigorous. If the logarithmic time derivative of the kinetic
energy is small however then $ s \sim \bar{T} t$ (where the bar denotes
an average over the region of interest). Otherwise a fully dynamical
formulation with time replacing $s$ in Equation (\ref{eq:L6}) would have
to be considered (P93; Cerruti-Sola \& Pettini 1995). 
The latter approach would also have to be used if the
curvature on $M$ is not predominantly negative. Also if the systems
we are comparing consist of different numbers of particles, $r_{\bf u}$
will have to be divided by $3N-1$ .

The definition leaves us the choice to compare 
different areas of the manifold of the
same dynamical system, or those of different systems. Also the averages
can either be static, or taken along a computed trajectory of the system.
Since the regions $R$ in the definition above can be taken as small as we
wish, the method is clearly local. This is possible
 because $r_{\bf u}$ is directly related to the local 
geometry of the manifold where a system lives and is not an
asymptotic quantity.  
Also, 
although the above formulation does not
allow explicitly for dissipative forces, these can be added as time
dependent perturbation to an open Hamiltonian system.

To actually calculate the value of $r_{\bf u}$, one contracts the Riemann
tensor (which for the metric (\ref{eq:L2}) is given in GS) and uses (\ref{eq:L5}) to
obtain
\begin{equation}
r_{u}=3A\frac{\left(W_{i}u^{i}\right)^{2}}{W^{2}}
-2A\frac{W_{ij}u^{i}u^{j}}{W}
-\left(A-\frac{1}{2} \right)\frac{\parallel \nabla W \parallel^{2}}{W^{3}}
-\frac{\nabla^{2}W}{2W^{2}} 
\label{eq:ru}
\end{equation}
with $A=\frac{3N-2}{4}$.
Here $W$ denotes $T=T(V)$ while $\nabla W$ and $\nabla^{2} W$ represent
its Cartesian gradient and Laplacian respectively. 
 From the metric~(\ref{eq:L2}) one can deduce that
$u^{i}=\frac{dx^{i}/dt}{\sqrt{2} W }$ and 
$\parallel {\bf u} \parallel=1$. 
For an $N$-body system,  the implied summation would be
over $i,j=1,3N$. Moreover, if the interactions proceed through the usual 
Newtonian law
with no direct impacts $\nabla^{2} W=0$. If we now label 
by $a$, $b$ and $c$ the particle numbers (which run from 1 to $N$)
and by $k$ and $l$ the three Cartesian coordinates of a particle, 
it is straightforward to obtain the following expressions for the derivatives of $W$
\begin{equation}
W_{i}=\frac{\partial W}{\partial x_{i}}=\frac{\partial W}{\partial
r^{k}_{a}}=- \sum_{c \neq a} \frac{r^{k}_{ac}}{r^{3}_{ac}},
\end{equation}
\begin{equation}
W_{ij}=\frac{\partial^{2} W}{\partial x_{i} \partial x_{j}}=
\frac{\partial^{2} W}{\partial r^{k}_{b} \partial r^{l}_{a}}=
\left[\frac{\delta_{kl}}{r^{3}_{ab}}- \frac{3r^{k}_{ab}
r^{l}_{ab}}{r^{5}_{ab}} \right],     \label{eq:ruw}
\end{equation}
if $a \neq b$ and
\begin{equation}
W_{ij}=\frac{\partial^{2} W}{\partial x_{i}  \partial x_{j}}=
\frac{\partial^{2} W}{\partial r^{k}_{b} \partial r^{l}_{a}}=
-\sum_{c \neq a} \left[\frac{\delta_{kl}}{r^{3}_{ac}}- \frac{3r^{k}_{ac}
r^{l}_{ac}}{r^{5}_{ac}} \right],
\end{equation}
if $a = b$. In these equations
\begin{equation}
r^{2}_{ab}=(r^{1}_{ab})^{2}+(r^{2}_{ab})^{2}+(r^{3}_{ab})^{2}
\end{equation}
and \hspace{0.1in}$r^{k}_{ab}=r^{k}_a -r^{k}_b$.

The practical procedure of implementing the criterion described above 
will therefore consist of using the position and velocities of particles
for the configurations of the system  under study
to obtain ${\bf u}$, $W$, and the quantities defined in~(\ref{eq:ruw}). These are substituted   into~(\ref{eq:ru})  
to find $r_{\bf u}$. 
In this way, one can calculate the Ricci curvature for various regions
of the configuration manifolds of different systems. If the
Ricci curvature is found to be predominantly negative, we then use
the above definition to classify systems according to their local stability
properties.

There are many advantages to the above setup.
For example, what is studied here is the normal deviation 
 due {\em random} perturbations
of trajectories 
with the same geodesic velocities ${\bf \parallel u \parallel}=1$ on $M$
and not the  deviation of temporal states in the direction of maximal growth.
Therefore what can be termed the `chaotic pendulum problem' is avoided.
In fact, it can be seen from~(\ref{eq:ru}) that all systems 
possessing only one  degree of freedom
 (those with $3N=1$) have an $r_{\bf u}=0$ at all times.  
In the terminology of classical stability theory it is said that the negativity
 of the Ricci curvature measures the orbital stability as
opposed to the more strict Liapunov stability (Pars 1965).
Because of these properties the Ricci curvature is unlikely to be 
negative for multidimensional integrable systems in virial equilibrium.
This assertion has been checked for the special case of the two body problem 
with circular orbits (where the virial relation is satisfied) but obviously
needs further investigation. For a system in a  full statistical quasi-steady state
a negative Ricci curvature  must mean that the system is chaotic and
mixing for all initial conditions since
$r_{\bf u}$ is constant for systems in a steady state. 
However, in general,
the negativity of the Ricci curvature on most of a system's trajectory does not guaranty that
it is chaotic 
but only that there is
a probability of this being the case (the probability increasing with 
the fraction of time spent in the negative region).


Other advantages of this method are that 
the integration of a large number of linearized equations
is avoided and space averages on compact manifolds can be compared to 
time averages (e.g., Casetti \& Pettini 1993), thus allowing one to check
for properties such as ergodicity for single particle orbits and
specially constructed  $N$-body systems for which $M$ is compact
(such as those consisting of softened particles enclosed
 enclosed in boxes: e.g., Lynden-Bell 1972).

\section{Applications}
\subsection{Some applications that take advantage of the geometric setting}
There are at least three applications that can make use of the 
geometric method described above.
\begin{enumerate}
\item
Studying the instability properties of individual orbits in fixed 
potentials with compact phase-spaces
(in this case the formulas in (\ref{eq:ruw}) will, of course, have to be
modified accordingly). 
\item
 Since $r_{\bf u}$  in Eq. (\ref{eq:ru})
depends only on functions that are given by sums over the particle positions
and velocities, it is constant (up to $1/\sqrt{N}$ fluctuations)
for systems in statistical equilibrium. 
In this case, it is  therefore possible  to replace time averages with phase-space
averages (that is different realization of same density and velocity fields).
This time however the averages are made over the full $6N$ phase-space. The
Ricci curvature therefore would give us a powerful tool of exploring that
space at and around equilibrium solutions. Important questions such as the degree
of chaos in a system, mixing time-scales, and the variation of these 
properties with particle numbers  can then be tackled in the $6N$
phase-space without making assumptions about the particle particle correlations.
 It is very important to compare such predictions with those
of single particle integrations in fixed potentials to evaluate the role
of discreteness in the evolution of gravitational systems. For as we shall
see in the next section, systems with large softening parameters appear
to have very different $6N$ phase-space structures compared to those
composed of point particles.  
\item
One can apply the method directly to the results of
$N$-body simulations. The formula for calculating the Ricci
curvature should be easily incorporated into large-$N$ codes such as the TREECODE. 
This would help in interpreting the results of
$N$-body simulations and lead to classification of galaxies according
to their dynamical instability properties. 
Moreover, one can then study  the direction of evolution of instability 
properties of realistic gravitational systems. 
For example it has been argued by Gerhard (1985) that elliptical galaxies
start from chaotic  states and evolve towards progressively more regular 
states which are then for long times indistinguishable from those of integrable
systems.
An alternative scenario is closer to the conventional dynamical interpretation
of statistical mechanics.
In this picture, a quasi-steady state is achieved when a system, although  
highly mixing,  keeps its macroscopic parameters constant 
and is stable against perturbations to its statistical properties. 
The point being that, if the instability 
is present for most initial conditions of interest (that is if we have 
constant negative $r_{\bf u}$ for long enough times), 
that would mean that the system is free to move in the region of interest 
and will tend towards a more probable state. This state would (by definition)
contain more microscopic states compatible with it and the system
would be free to move between them.
  This situation is more compatible with the
interpretation of violent relaxation as leading to most probable
end states compatible with a set of constraints (Saslaw 1985)
since regular states cannot be very probable 
because they lie on $3N$ subspaces of the $6N-c$ 
dimensional subset of the phase-space where all
possible states live ($c=10$ stands for the number of classical
integrals used in the reduction of the 
$N$-body problem, e.g., Whittaker 1937).
\end{enumerate}

\subsection{Specific application and model parameters}
  
 As a first test we  apply the method described  above to small
$N$-body systems of 231 point particles. The small number enables us to
integrate the equations of motion with high precision, on the other hand the number
should be sufficient for the results to have some statistical
significance. It is  essential to check if the 
Ricci curvature method makes adequate predictions about the evolution
of gravitational systems under these controlled conditions before 
applying it to realistic galaxy models where a host of auxiliary problems
will arise. The small numbers however makes it harder to distinguish
clearly between the predictions of the  Ricci method and those of two body
relaxation theory. A detailed
comparison (including dependence of the results on $N$) is better left to another study.
 
We choose initial conditions in which the particles are arrayed
into two sheets. An upper one with 11 lines consisting of 
11 particles each and a lower one composed of 11 lines with 10
particles each. The lines of the upper and lower sheets are positioned
in such a way that a line in the lower sheet lies at half the distance
(in the plane of the sheet) between two lines in the upper sheet, so
as to avoid direct contact of particles from two different sheets when the system evolves. The separation between the two sheets is taken to
be equal to half the separation between lines in the same sheet (see the
$t=0$ snapshots in Fig. 3 for example). Such  configurations are  artificial
and are therefore likely to quickly and visibly evolve, 
hence saving us the trouble of long time integrations.

We use units in which the gravitational constant is unity, 
the masses of all particles are also taken as unity. 
Time in these units will be referred to as ``physical time''.
Three scales are used to determine the separation between adjacent particles
$d =100 \times Scale$,
where $Scale$ takes either the value of 1, 10.8 or 100. In what we may call the ``main models'' we give the initial configurations a rigid
body rotation corresponding to an angular momentum of 44275 units around the Z-axis (with zero initial velocities in the Z direction), or 
a random number generator is used to fix the X-Y velocities which
give rise to a small angular momentum of 62.61 units. The three cases of
$Scale=(1,10.8,100)$ correspond to energies of $(-41.471,-5.705,-0.635)$
with corresponding initial virial ratios of $(0.695,0.0644,0.00695)$ respectively. Alternatively, some runs are started with a fixed initial
virial ratio of 1 and with the same energies as above. The angular momentum is adjusted accordingly. We shall call these the ``equilibrium
models'' although they do not start from a detailed dynamical equilibrium.

The integrations are performed using a variable order variable step size
Adams method as implemented in the NAG routine D02CBF using a tolerance of $10^{-13}$. The energy is accordingly conserved
to better than 10 digits
(usually 12)  for a few dynamical times. This accuracy is 
necessary for a first numerical test to eliminate the factor of serious
numerical error from the interpretations of the results. For the same 
reason we integrate the equations for a relatively short time of about
four and a half dynamical times 
 ($\tau_{D}=$(mean density)$^{-1/2}$),
 $4.24 \tau_{D} \sim \tau_{b}$) corresponding to about 2200 time units
for the $scale=1$ case 
(in a few cases we have performed longer time integrations). 
The un-softened force law is used, and softening
is only introduced to study its effect.

\section{Results of numerical experiments}
\subsection{Averaging}

\begin{figure*}
\begin{flushleft}
\epsfig{file=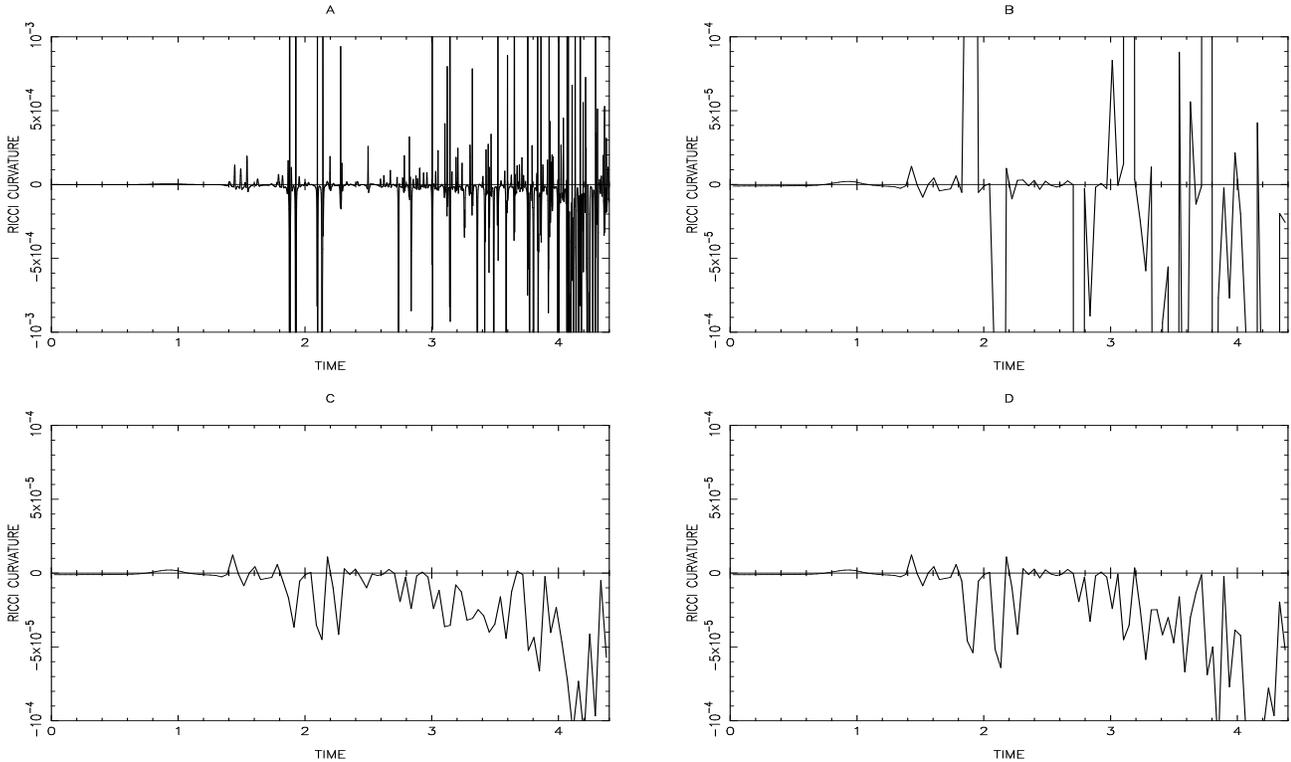,width=17.0cm,height=10.0cm,angle=-90}
\end{flushleft}
\caption{\label{average}
Ricci curvature time series for rotating ``main model'' $Scale=1$
system: {\bf A} Unaveraged time series, {\bf B} with averages taken
over a hundred intervals, {\bf C} as in {\bf B} but with a ``filter'' of $2 \times 10^{-4}$,
{\bf D} as in {\bf B} but with a filter of $4\times 10^{-4}$ 
and $10 \times 10^{-4}$ (dotted line)}
\end{figure*}

In order that our criterion may be useful, the Ricci curvature must 
remain mostly negative throughout the evolution of a given system.
This condition cannot be generally satisfied  because
$r_{\bf u}$, as given by~(\ref{eq:ru}), is singular as 
$r_{ab} \rightarrow 0$ for any $a$ and $b$. The time for which
$r_{ab}$ is near zero must also be small compared to the dynamical
time-scale of the system (a bounded circular orbit of two particles of separation 0.1  has a period of $\sim 2 \times 10^{-4} \tau_{D}$ with $\tau_{D}$  the dynamical time of the $Scale=1$
main model). Therefore, during an encounter of two or more point masses,
their contribution will fluctuate violently and dominate $r_{\bf u}$, this contribution
will have a  positive average since for a time it mimics that of a two body system. 
Thus we expect $r_{\bf u}$ to be a very ``bumpy'' function of time 
for $N$-body systems of point particles.  This is indeed the case, as we can see from Fig. 1A, where the Ricci curvature is plotted as a function of dynamical time for the rotating main model with $Scale=1$  (see
section 3.2). We will therefore have to try to eliminate this bumpiness
 in the hope that the residual curvature is negative if the system is mixing.
The  approach we will use here hinges on a theorem by Bogoliubov (P93 and the
references therein) which states that for equations of the type~(\ref{eq:L6})
averaged over small time-scales, the solution is similar to that of the 
original equation averaged over that time-scale if one excludes 
the possibility of parametric instability (e.g., ARN) and  direct singularities; 
this is the method we try in this study. 
We will not attempt here to find out what
the theoretically optimum way of doing that should be, but will proceed empirically.  As a first approach we will just average over a time-scale containing many ``bumps'' but still small compared to the total integration time. 

Fig. 1B shows the time series in Fig. 1A averaged over steps of $(1/100) \times
 4.4 \tau_{D}$ (corresponding to 22 data points in Fig 1A)
we see that now the series becomes much more regular. To make further 
progress we remove the contributions from the most extreme peaks 
which can dominate the average giving misleading results (these peaks
represent points nearer to singularities). 
It is clear from Fig. 1A that peaks with absolute 
values larger than $2-4 \times 10^{-4}$ are rare and 
isolated, one therefore is tempted to filter the results 
so as not to include in the calculation of the average any peaks larger 
than a threshold of this order of magnitude. Fig. 1C shows 
the behaviour of $r_{\bf u}$ when a threshold of $2 \times 10^{-4}$
is taken. The striking result is the apparent regularity and negativity of the Ricci curvature in this case.

It is now important to check that these results do not 
sensitively depend on the values of the threshold used or the 
averaging interval taken. Fig. 1D shows the behaviour of $r_{\bf u}$ when a threshold of 
$4 \times 10^{-4}$  and $10 \times 10^{-4}$ (instead of $2 \times 10^{-4}$) 
is used. The results clearly appear to be qualitatively similar. Indeed, 
it has been found 
that the results start to become radically 
different only for a threshold $\sim 20 \times 10^{-4}$. 
Unless stated otherwise, it is implied  that it has been
checked that the results do not sensitively depend on the threshold taken and that a value of $2 \times 10^{-4}$ has been adopted. 
It has also been checked that the results do not sensitively depend on the
size of the
subdivision intervals (as long as they are not too small of course).
We will take averages over a hundred subdivisions of the integration 
interval unless otherwise stated.
\subsection{Rotating versus Non-rotating models}

Rotation, plays a central role in stellar dynamics and is one of
the parameters that vary  along the Hubble sequence.
 Intuitively, systems where random motion 
dominates are expected to mix faster (at least initially) than 
ones where ordered rotational motion does. Observations of elliptical
galaxies also show them to be highly mixed and relatively (in comparison
with spirals)
relaxed objects so that an effective mixing mechanism must have been at
work at some stage (and could still be).
Our first application will therefore be 
the comparison of a rapidly rotating system to one without (almost) 
any rotation.
 
\begin{figure}
\begin{flushleft}
\epsfig{file= 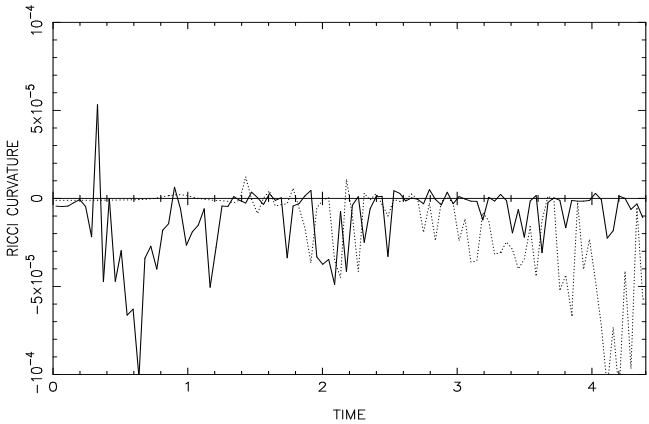,width=8.5cm,height=5.5cm,angle=-90}
\end{flushleft}
\caption{\label{ranrot}
  Ricci curvature time series for main model with random initial $x$-$y$ velocities
and $scale=1$ (solid line) averaged as in the plot in  Fig.~\ref{average}~{\b C} (reproduced here by the
dotted line)
 }
\end{figure}

In Fig. 2
we have plotted the time series of $r_{\bf u}$ 
for the two main models 
with $Scale=1$. Shown by the solid line is the series corresponding to 
the random initial velocities, while the dotted line represents the 
evolution of $r_{\bf u}$ for the rotating case. 
Initially, the random system is much more unstable as expected, while later
on it starts evolving on a longer time-scale (characterized by the small
$r_{\bf u}$)  while the reverse is true for the rotating system.

\begin{figure}
\begin{flushleft}
\epsfig{file= 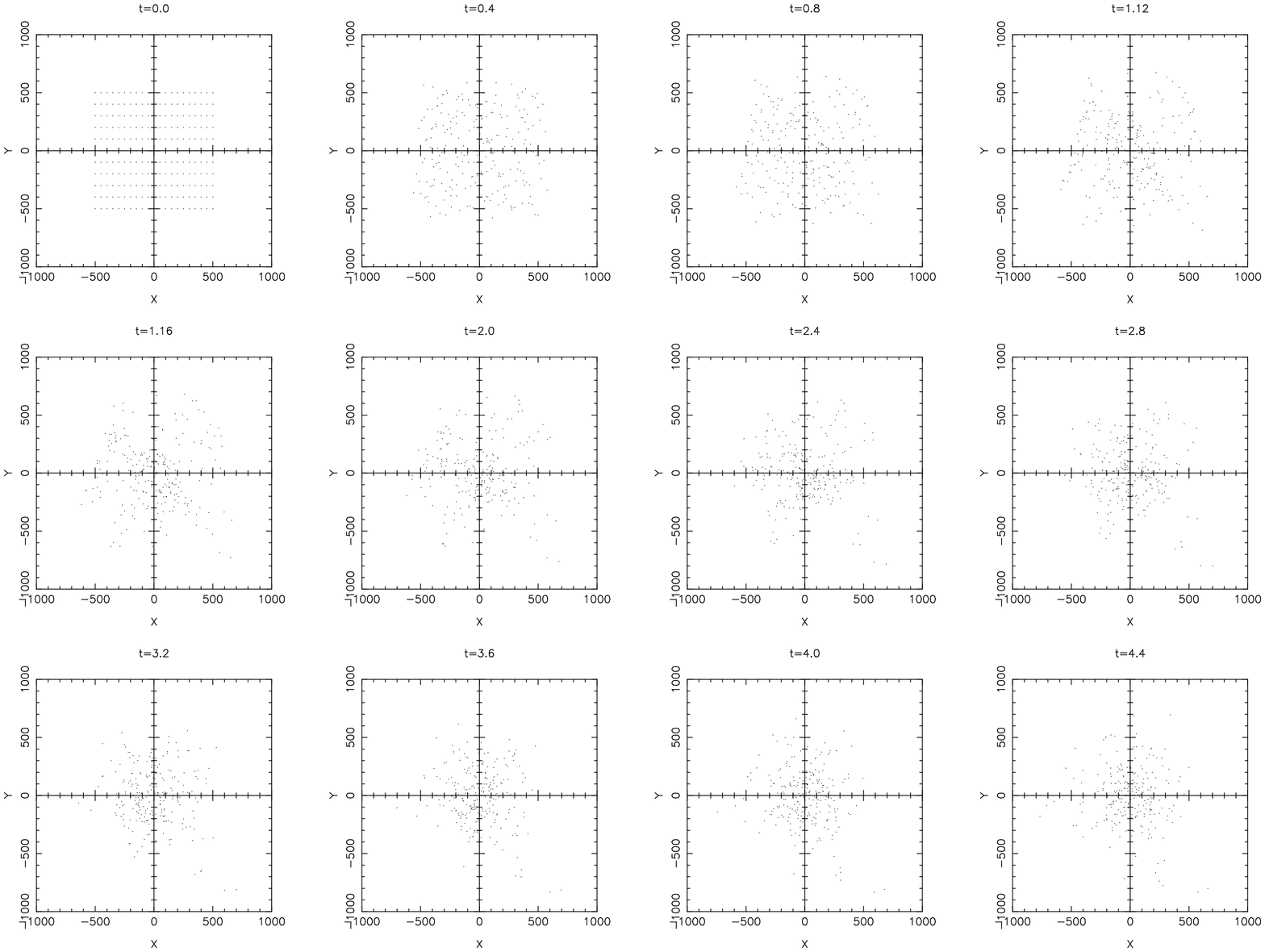,width=8.7cm,height=7.2cm,angle=-90}

\vspace{0.10in}

\epsfig{file=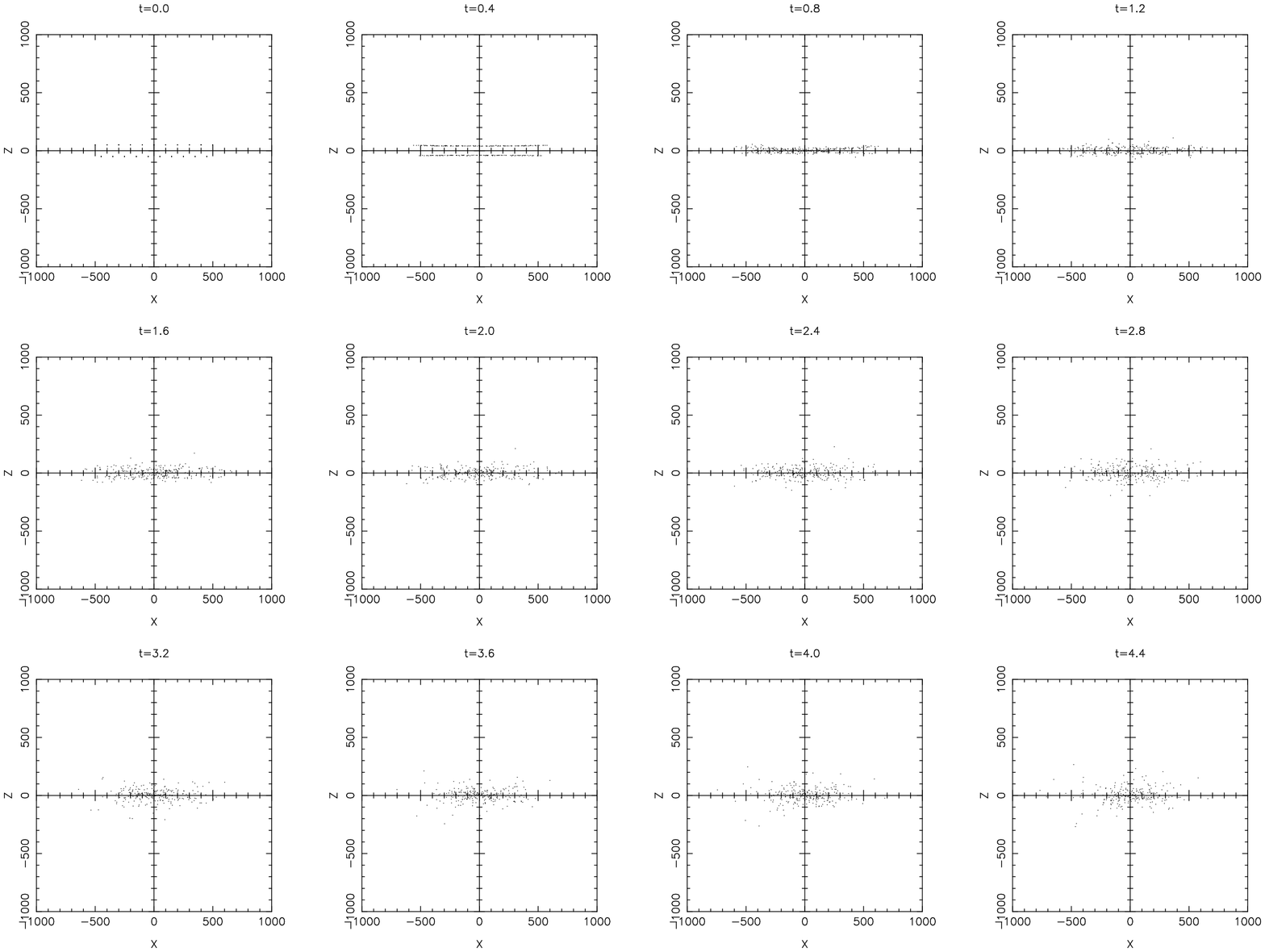,width=8.7cm,height=7.2cm,angle=-90}
\end{flushleft}
\caption{\label{mod1rxy}
Spatial evolution of the model corresponding to solid line plot in  Fig.~\ref{ranrot} }
\end{figure}

\begin{figure}[h]
\begin{flushleft}
\epsfig{file=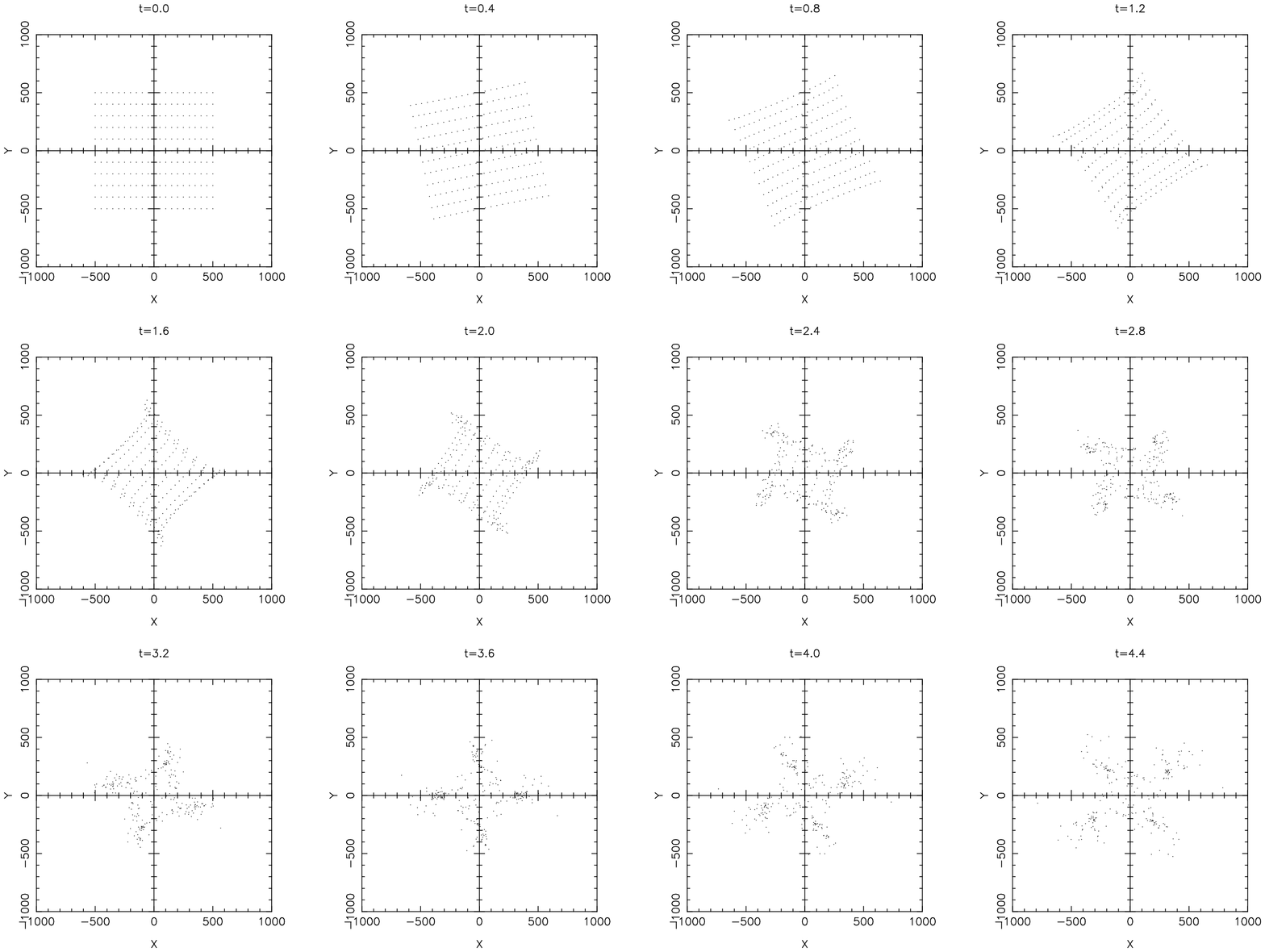,width=8.7cm,height=7.2cm,angle=-90}

\vspace{0.10in}

\epsfig{file=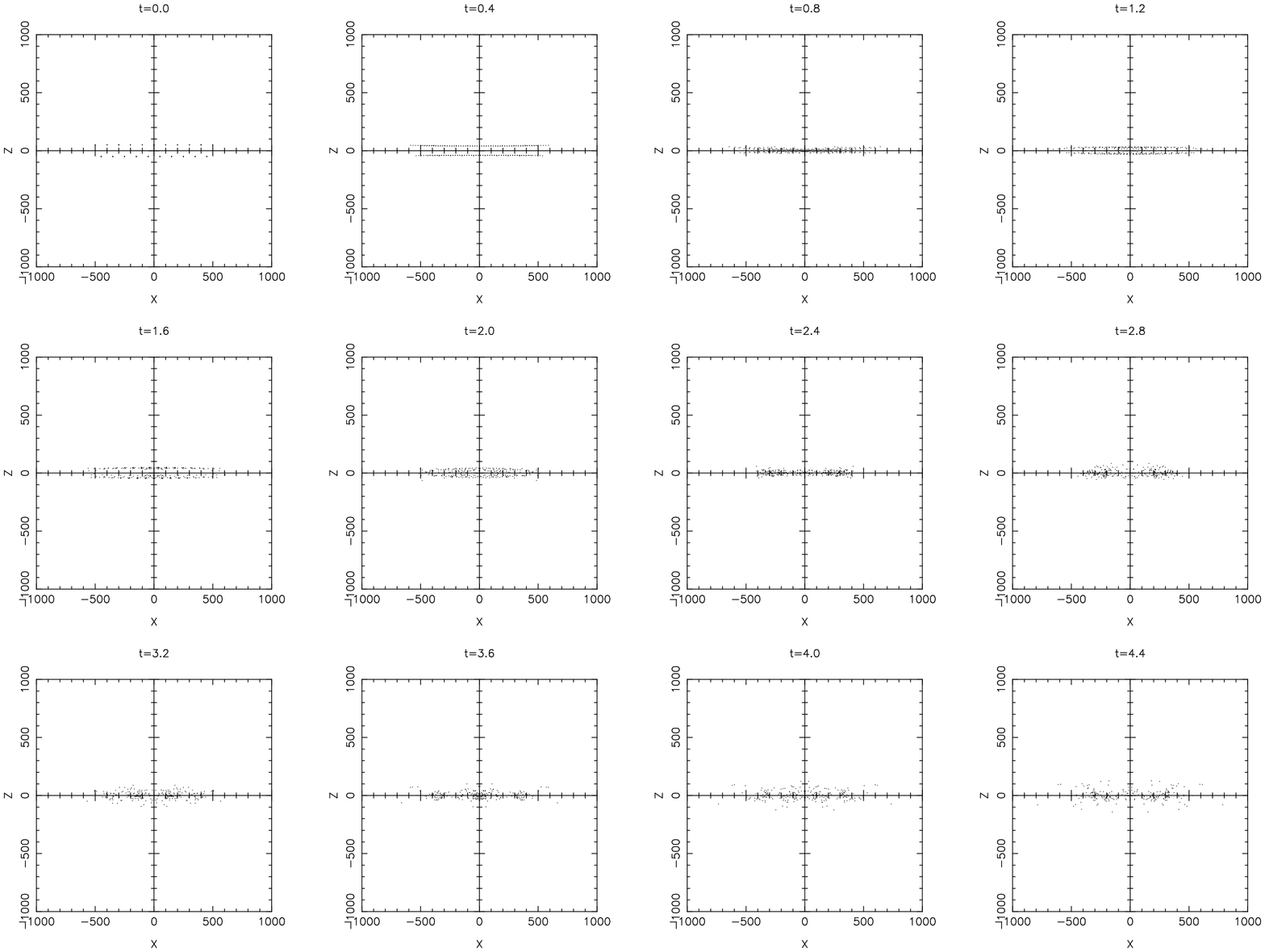,width=8.7cm,height=7.2cm,angle=-90}
\end{flushleft}
\caption{\label{mod1xy}
Spatial evolution of the model corresponding to the dotted line plot in Fig~\ref{mod1rxy}
 }
\end{figure}

Explanation of this behaviour of $r_{\bf u}$ may be obtained by 
looking at the spatial evolution of the two systems. The random system 
evolves much faster initially and quickly loses memory of its 
initial configuration in the $x-y$ projection  while becoming 
diffuse for times beyond $t=1.2-1.6$ and starts to puff up in 
the $z$-direction (Fig. 3) thus evolving to a more isotropic state. 
However since it is becoming more diffuse it now takes longer to evolve 
because its density is decreasing. 

On the other hand, 
the rotating system keeps its shape almost
perfectly intact until $t=2$ when a distinctive 4 armed clustering pattern (Fig. 4)
starts to appear accompanied by the formation of high density areas. 
This explains the large negative values of the   Ricci curvature which in this case
represents
the collective instability. Since geodesic instability, as stressed earlier, 
is independent of the particular evolutionary phenomenon in 
question, this type of relaxation is also included in the description.
When one interprets relaxational phenomena in terms of  thermodynamic 
``more probable states'' therefore, it should be clear that those types of
instabilities could also occur. If they 
do not lead to complete detachment of a system into components however, 
the picture that emerges, is that while these  phenomena 
may be important for systems for which the classical theory holds,  
they cannot characterize the long term evolution of a 
generic system 
and that these inhomogeneities themselves produce 
chaos and evolution towards more probable 
states under small dissipative or conservative perturbations 
(Pfenniger \& Norman 1990; Hasan et al. 1993;
Pfenniger \& Friedli 1991; Friedli \& Benz 1993; Zhang 1996). 
They are therefore not typical properties 
of gravitational
systems but transient ones. That, roughly speaking,
means that regions of phase-space where
this type of instability occurs should be small compared with those where
the instability leading to more isotropic states arises.

We now calculate a rough evolutionary time-scale corresponding to   
 the plots in Fig 2. We take it to be {\em the average 
exponentiation rate for small random
perturbations normal to the phase-space path of the $N$-body systems}. From the
considerations of Section 2.2 this is given by 
\begin{equation}
\tau_{e} \sim \left( \frac{3N}{-2 \bar{r}_{\bf u}} \right)^{1/2} \times \frac{1}{\bar{W}},     
\label{dyn}
\end{equation}
where a bar denotes a time average. Now, $r_{\bf u}$ averages to about
$-10^{-5}$ over the whole interval (for both systems) while $W$ starts at a
value of 22 but then rises to about 42, therefore we take a value of 30
as a rough mean. This gives a mean exponentiation time-scale of about 200 
units---or $0.4 \tau_{D}$ over this period of time (0 to $4.4 \tau_{D}$).

Now, what does this time-scale mean? To answer this question it is important
to remember that while the exponential solutions are local and arise from
a linearization of the dynamics, this instability exists everywhere because
of the negativity of the average of the Ricci curvature. Moreover, the instability
is that of the flow in  the full $6N$ phase-space and not just particle orbits. 
Under these conditions, the divergence  will probably lead to mixing 
and filamentation which  will radically alter the local structure of the 
$6N$ phase-space (as mapped by an arbitrary initial partition of that space).
Up to resolution $\epsilon$, the phase-space will be completely modified after
a time apparently given by (e.g., GS):
\begin{equation}
\tau_{\epsilon}= \ln \epsilon^{-1} \tau_{e}. 
\end{equation}
For a numerical application in double precision the smallest resolution available
 is $\epsilon \sim 10^{-14}$ (incidentally, this number is not too different from
the ratio of the size a star to that of a galaxy)
so that after a time 
\begin{equation}
\tau_{\epsilon} \sim  32 \tau_{e}
\label{eps}
\end{equation}
there is complete change in the system's phase-space for all practical purposes.
For an ensemble of trajectories having the  same energy and
integrated independently in a fixed potential,
Merritt \& Valluri (1996) found that complete mixing did indeed occur over a number
of exponentiation time-scales (measured by Liapunov exponents in their case)
comparable to that given  by (\ref{eps}).

 Examples of systems that were proved to behave in this manner include
the general class of C-systems (e.g., Arnold's famous cat map: LL). Although we cannot claim
that the process is as rigorous in our case, with its non-compact phase-space and sign 
indefiniteness of the 2d curvatures,
 we will follow the assumption that
it is at least qualitatively similar (the negativity of the  averaged Ricci
curvature, the results of GS and Kandrup 1990a,1990b suggest
that this may be so). In that case, we have complete modifications of our systems on 
a time-scale of about 13 dynamical times. Obviously, our two examples have
evolved significantly over a time-scale of this order of magnitude --- albeit
each in different ways and non-uniformly in time. 
We shall see in Section 4.4 that on a time-scale of $12 \tau_{D}$
there is indeed complete modification for one similar system.

Finally we note that formula (\ref{for 2})  predicts an exponential
time-scale of about $1.1 \tau_{D}$ which is somewhat larger than the one
derived from the Ricci curvature but within reasonable bounds
considering it is only an order of magnitude estimate derived for
infinite systems and does not take into account effects due to the large scale
gravitational field (see Kandrup 1989,1990b)
that lead to the clustering in the rotating case for example. 
  
\subsection{Varying the energy and virial ratio}

We now consider systems with lower densities while keeping the angular 
momefinger ntum constant. For systems initially in a state of collapse ($vir<1$)
this amounts to varying the energy --- making it larger. For these systems, the initial virial ratios (and hence the velocities)
is   very small, therefore the results for the rotating systems are
 more or less similar to those for systems starting with random X-Y velocities.
We will therefore focus on the latter type only. 

\begin{figure}
\begin{flushleft}
\epsfig{file=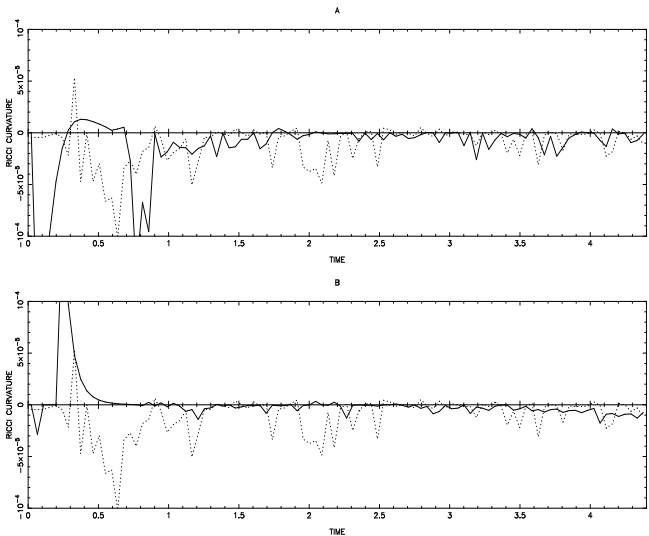,width=8.5cm,height=7.0cm,angle=-90}
\end{flushleft}
\caption{\label{ranvil}
 Ricci curvature time series for random main model systems with: {\bf A}
$Scale=10.8$, {\bf B} $Scale=100$ averaged as the plot in  Fig.~\ref{average}{\bf C} (reproduced here
by the dotted lines)}
\end{figure}

In 
Fig. 5 we compare the Ricci time series for initial conditions with $Scale=10.8$
(Fig. 5A) and $Scale=100$ (Fig. 5B) (shown by the solid lines) 
with those of the $Scale=1$ (dotted line). 
Clearly, there is a trend towards larger time-scales of evolution for the less dense
systems, especially if one takes into account that the ratio of the kinetic
energy $W$ at virial equilibrium
 of the $Scale=1$ system to that of the $Scale=10.8$ and the $Scale=100$ systems is 
7.3 and 65.9 respectively. These results may seem counterintuitive since one 
knows that for initial conditions far from dynamical equilibrium there is a
well known
instability at work --- namely violent relaxation. The confusion is resolved
  however when one notices that the dynamical time for these two systems is much larger 
than the original --- being 35.49 and 1000 times larger. Thus, based on the above 
considerations, we expect a system characterized by the Ricci curvature 
shown by the solid line in Fig. 5B to evolve extremely slowly in ``physical''
time units but relatively fast, compared to the one shown by the dotted line, 
in terms of dynamical time. The fact that the kinetic energy varies rapidly along the 
motion prevents us however from calculating any precise estimates.

\begin{figure}
\begin{flushleft}
\epsfig{file=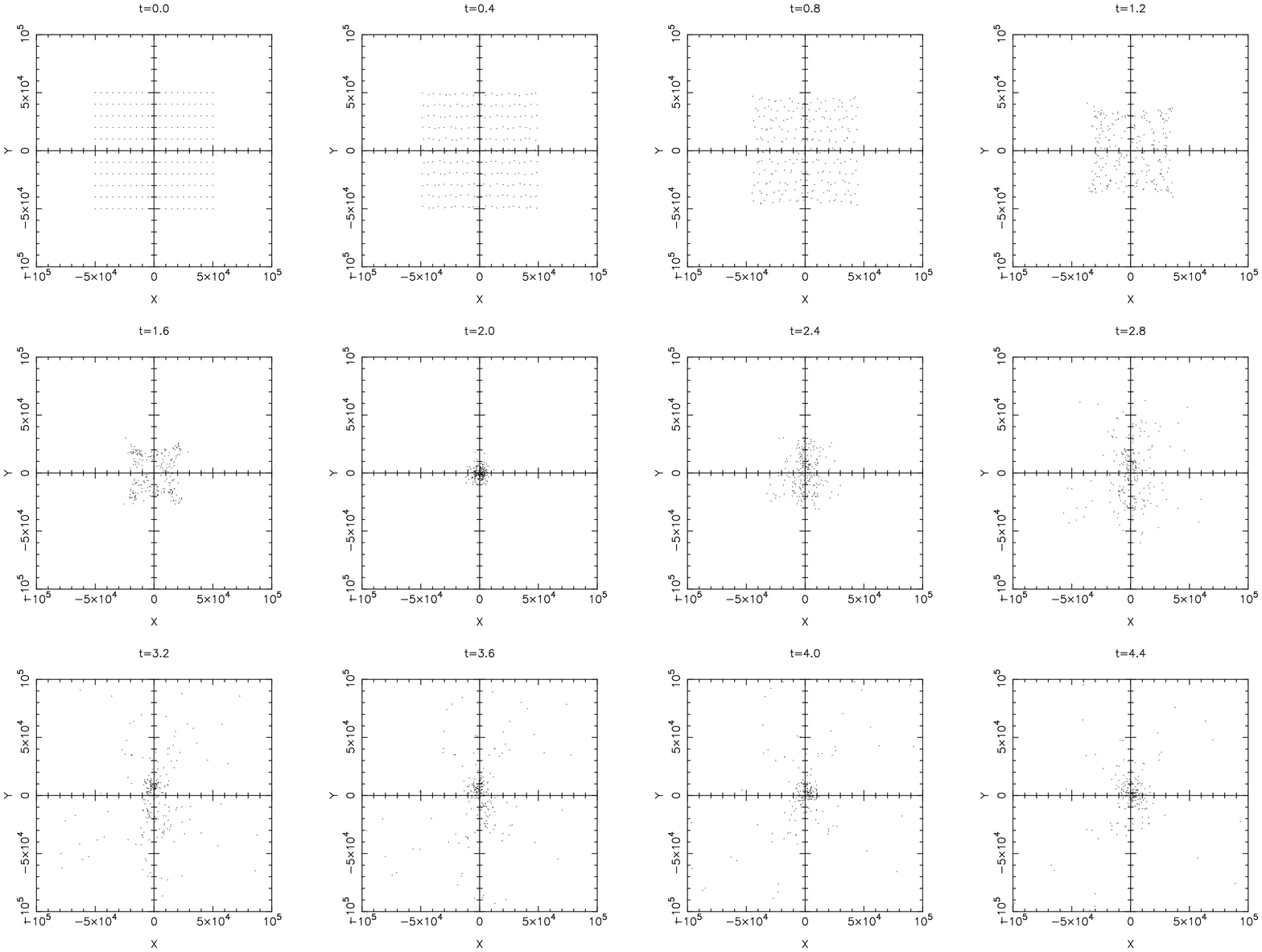,width=8.7cm,height=7.2cm,angle=-90}

\vspace{0.10in}

\epsfig{file=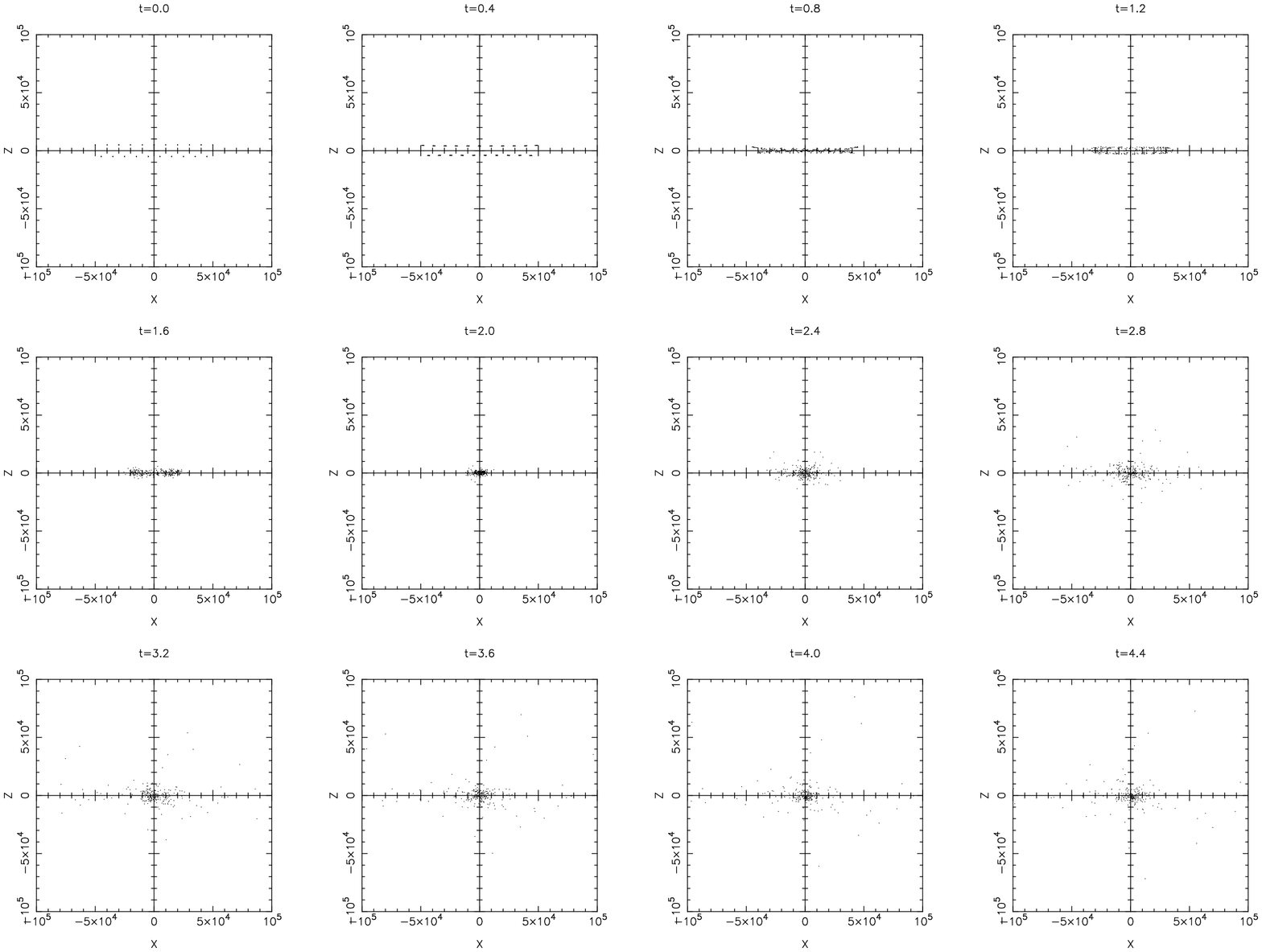,width=8.7cm,height=7.2cm,angle=-90}
\end{flushleft}
\caption{\label{mod6rxy}
Spatial  evolution of the model corresponding to the solid line 
plot in  Fig.~\ref{ranvil}~{\bf B} }
\end{figure} 

 By looking at Fig. 6 one can clearly see that this
is the case. With every interval between ``snapshots'' in this figure 
corresponding to a thousand intervals in Fig. 3 one sees that the $Scale=100$
system evolves much more slowly in physical time. In terms of intrinsic 
dynamical times however, it is clear that it evolves much faster than the 
$Scale=1$ system; having quickly lost all trace of its initial spatial 
configuration it then appears to be evolving towards a state characterized by an increasingly tight core and
diffuse ``halo''. At this stage one notices, 
the almost constant (drifting slowly towards lower) negative values   of $r_{\bf u}$ in fig 5B (solid
 line) beyond $t \sim 3$, showing
that the system is becoming more and more mixing, in agreement with discussion 
in section 3.1.
 
\subsection{Systems starting in virial equilibrium}
We now keep the  energies constant and increase the angular momenta of the
rotating systems 
in such a way as to have virial equilibrium in the initial state.
This will be accompanied by
contraction in the scale of the system. For example, the $Scale=100$
equilibrium models will have an inter line spacing of about half
that of the main models with the same energy. The systems with random $x-y$
velocities start from the same spatial configurations as the rotating ones 
and with the absolute values of the velocities rescaled accordingly.

\begin{figure}
\begin{flushleft}
\epsfig{file=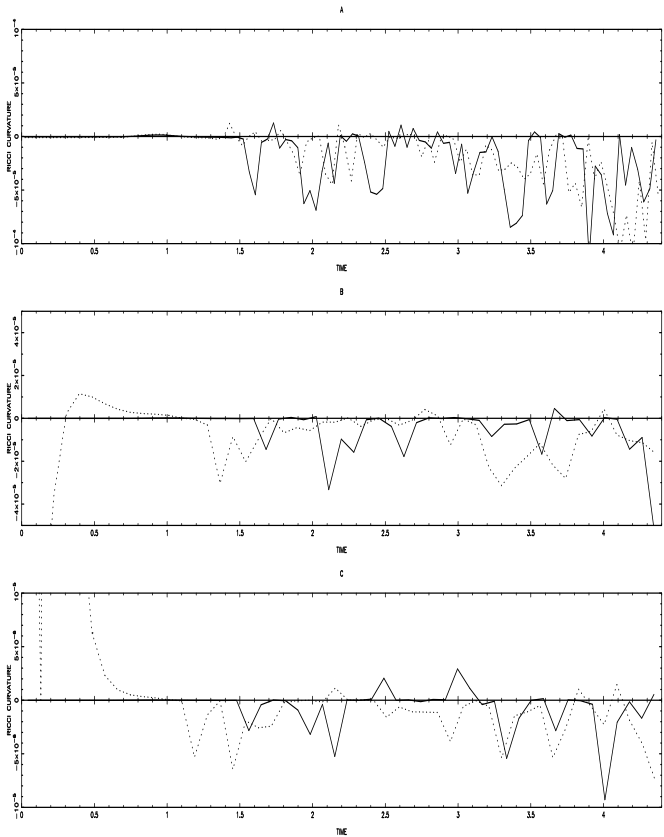,width=8.7cm,height=11.0cm,angle=-90}
\end{flushleft}
\caption{\label{virplot}
 Ricci curvature time series for random main model systems with: {\bf A}
$Scale=10.8$, {\bf B} $Scale=100$ averaged as the plot in 
 Fig.~\ref{average}{\bf C} (reproduced here
by the dotted lines)
Ricci curvature time series of the rotating  equilibrium models (solid lines)
compared with the main (collapsing) models: {\bf A} $Scale=1$, {\bf B} $Scale=10.8$,
{\bf C} $Scale=100$. In {\bf B} and {\bf C} averages are taken over 50 time intervals (instead of a 100)
 }
\end{figure}

For a system in a steady state we expect $r_{\bf u}$ to be constant in time.
If, in addition, the motion is regular (e.g., our sheets rotate rigidly)
we expect $r_{\bf u}$ to be very near zero. Nevertheless, due to the 
differential rotation which is accompanied by collective instability and 
clustering, this state of affairs cannot last too long. Similarly, due to the
instability of the initial state
in the random case, the early equilibrium is lost. 
These considerations are confirmed by plots of the Ricci curvature (Fig. 7)
where values for the equilibrium cases (solid lines) are compared with 
 those of main models of the same energies (dotted lines). Although
the curvature is still negative even at early times, the evolution
time-scale given by Eq. (\ref{eps})  is now about 70 dynamical times.
It also is immediately apparent that the more distant the original system was
from virial equilibrium the bigger the difference of its initial behaviour
from its equilibrium counterpart. For example, for the $Scale=1$ system (Fig 7A)
the dotted and solid lines are almost indistinguishable, while for the
other two systems the contrast in the early evolution is much clearer. 
Later on however, non-equilibrium systems quickly tend towards virial equilibrium, making the contrast weaker.
 (The reason for the positive
bumps in the time evolution of the curvature for these models
is that, despite the fact that we average
over 50 time intervals, the irregularities cannot be 
suppressed because the filtering is too large the fluctuations
now are much smaller because of the lower density.)  
It is clear from these graphs that, because
the dynamical time of the equilibrium systems is smaller than the 
collapsing ones (about a third for the $Scale=100$ models), the equilibrium
models actually
evolve slower than the collapsing systems in terms of intrinsic dynamical
times throughout most of the evolution. From that, and from the results of the
previous subsection, we conclude that systems starting from a state of collapse
are possibly more unstable than ones starting near virial equilibrium. The difference does not  however  appear to be very large.

\begin{figure}
\begin{flushleft}
\epsfig{file=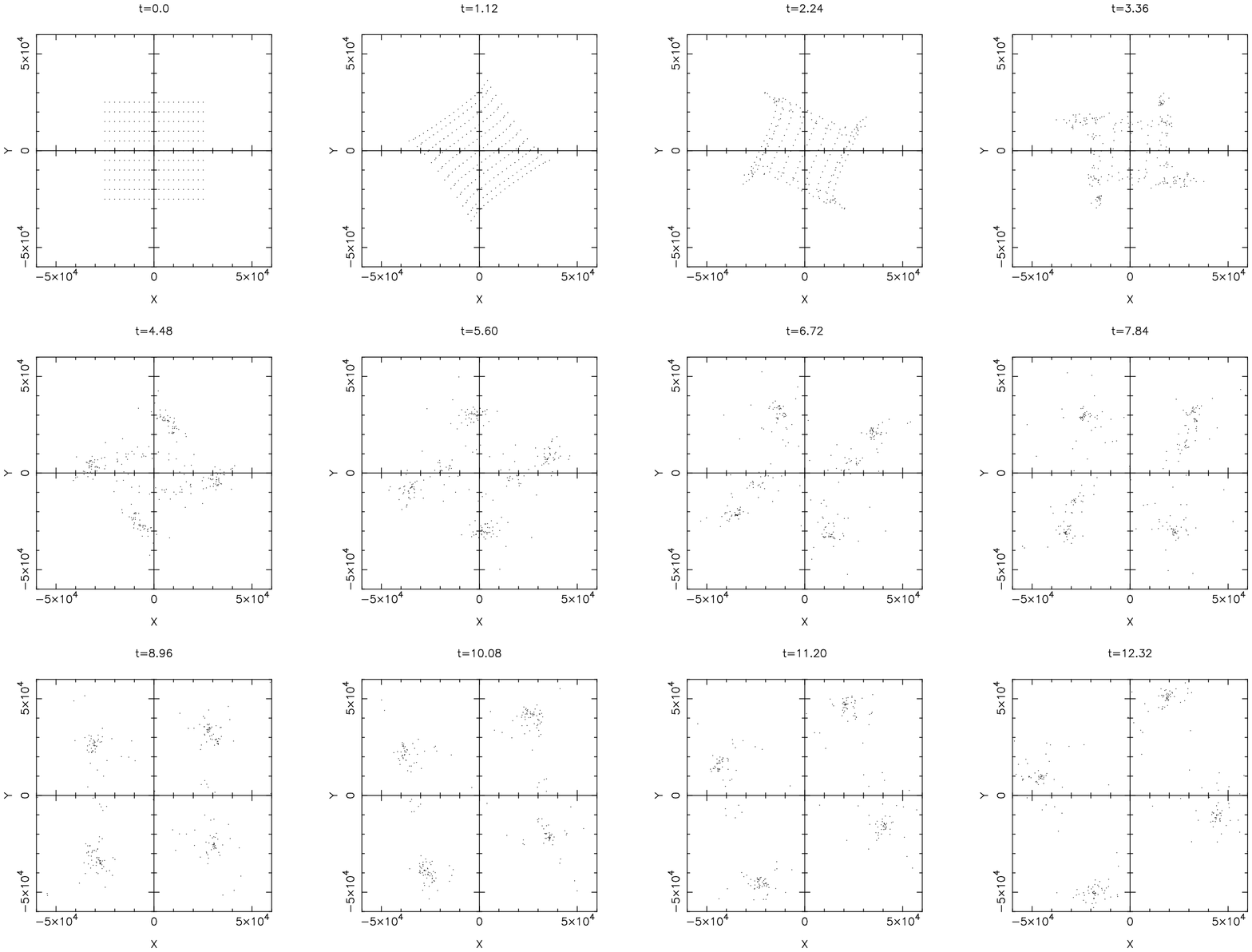,width=8.7cm,height=7.2cm,angle=-90}

\vspace{0.1in}

\epsfig{file=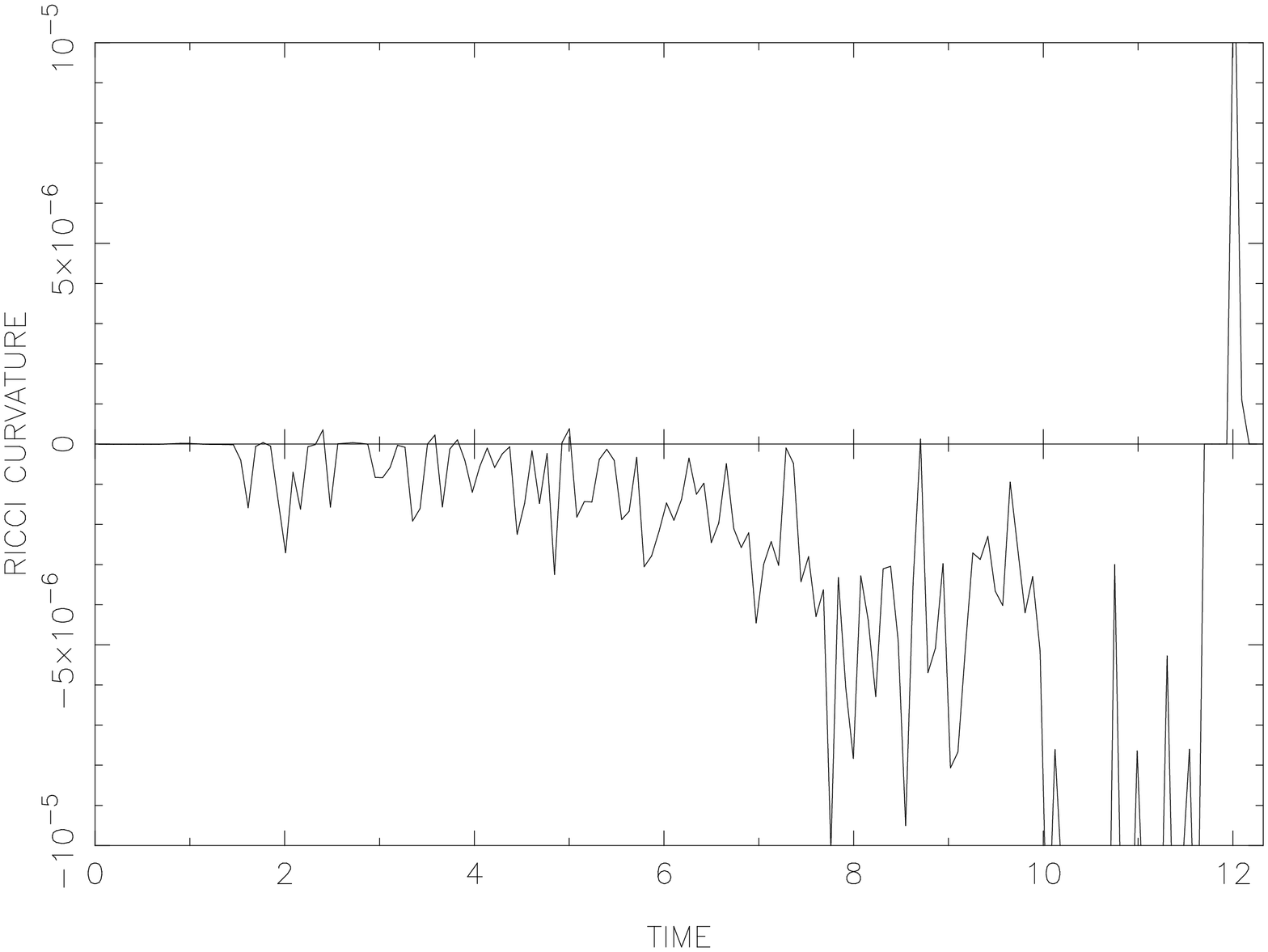,width=8.7cm,height=2.5cm,angle=-90}
\end{flushleft}
\caption{\label{mod7rxy}
  Longer time behaviour of the Ricci curvature and the corresponding $x$-$y$
projection of the spatial evolution for the system  corresponding to the 
solid line plot in 
 Fig. 7C. Filtering is
taken equal to $2 \times 10^{-5}$ and averaging is done over 150 intervals.
}
\end{figure}

The $x-y$ spatial  evolution in 
time of the  rotating $Scale=100$ equilibrium model is shown in
Fig. 8 where the integration was carried over a time interval corresponding 
to the physical time of integration of its main model counterpart with same
energy. It is clear that by $t=4.48$ the clustering is at least as pronounced 
as in Fig. 4 at $t=4.4$ (a similar comparison can be made with the corresponding 
equilibrium model which behaves in a very similar manner). This is not surprising
given the scale free nature of gravitational interactions.
 It is therefore
interesting to see what the rough time-scales derived from the Ricci curvature 
might tell us about this apparent coincidence. Let us label the quantities 
relating to the $Scale=1$ equilibrium 
initial state by indices 1 and those relating to
the $Scale=100$ equilibrium system 
 by 100 . The ratios of the two evolutionary time-scale are 
then
\begin{equation}
\frac{\tau_{100}}{\tau_{1}}
=\sqrt{ \frac{ \bar{r}_{{\bf u}1} } { \bar{r}_{{\bf u}100} } }
\frac{\bar{W}_{1}}{\bar{W}_{100}}   
\frac{\tau_{D1}}{\tau_{D100}}. 
\end{equation}
Both systems are always near virial equilibrium throughout 
the evolution therefore  the W's are constrained to be near values given
by $W=-E$ so that $\bar{W}_{1} \sim 41.53$ and $\bar{W}_{100} \sim 0.63$, the 
averages $\bar{r}_{\bf u1}$ and $\bar{r}_{\bf u100}$ are calculated over
the interval $t=0$ to $t=4.4$ to be $\bar{r}_{{\bf u}1}=-1.6 \times 10^{-5}$
 and $\bar{r}_{{\bf u}100} =-5.8 \times 10^{-7}$ . Finally  
\begin{equation}
\frac{\tau_{D100}}{\tau_{D1}}=\left(\frac{d_{100}}{d_{1}} \right)^{3/2}=(50.17)^{3/2}=355.41,
\end{equation}
where $d_{1}$ and $d_{100}$ are the initial separations of adjacent particles
on the sheets.
Inserting these numbers we get
\begin{equation}
\frac{\tau_{100}}{\tau_{1}}=0.97.
\end{equation}
Now, considering that the systems differ radically in energy, angular momentum 
and dynamical time-scales this result is rather impressive---especially if 
one recalls the rough way in which we have averaged and filtered the data.
It may be useful to note that Eq.  (\ref{for 2})  predicts the same
behaviour, we have
\begin{equation}
\tau_{gs} \sim \frac{W^{1/2}}{n^{2/3}}\sim
\frac{W^{1/2}}{n^{1/6}}\tau_{D} \sim \left(Scale \times W \right)^{1/2}
\end {equation}
which gives \begin{equation}
\frac{\tau_{gs100}}{\tau_{gs1}}=0.87.
\end{equation}

Looking at the longer time behaviour in Fig. 8 we see that
the clustering continues leading to almost total separation
of the components. The corresponding behaviour of $r_{\bf u}$ 
is also shown. As would be expected, the clustering gives rise to
shorter evolutionary time-scales. Three 
characteristic time-scales can be isolated: 
from $t=0$ to $t=7$ the system can still be considered to be one part
and is connected until $t \sim 10$ but then detaches after that. These 
stages are clear from Fig. 8 . Thus there is complete modification of the 
initial state after about 12 dynamical times as expected from Eq. (\ref{eps}).  
To illustrate how the results depend on
filtering, we have chosen a filtering here of $2 \times 10^{-5}$
instead of the value $ 2 \times 10^{-4}$ usually used. Hence the
smoothness of the time series up to $t=4.4$ in comparison with
the one in Fig. 7C. The positive bump  on the right  is due to the 
filtering becoming too small relative to the absolute value of $r_{\bf u}$
so that the value in this last interval has little statistical significance.

Similar results are obtained for the cases with random initial
$x-y$ velocities.  Here, the formula for two body
relaxation for a Maxwellian distribution 
(BT formula 8.71 which is more or less appropriate
in this case) gives a ratio of $\tau_{100}/\tau{1}=0.87^{3}=0.66$ for these two
cases. 

\subsection{The effect of softening}


\begin{figure}
\begin{flushleft}
\epsfig{file=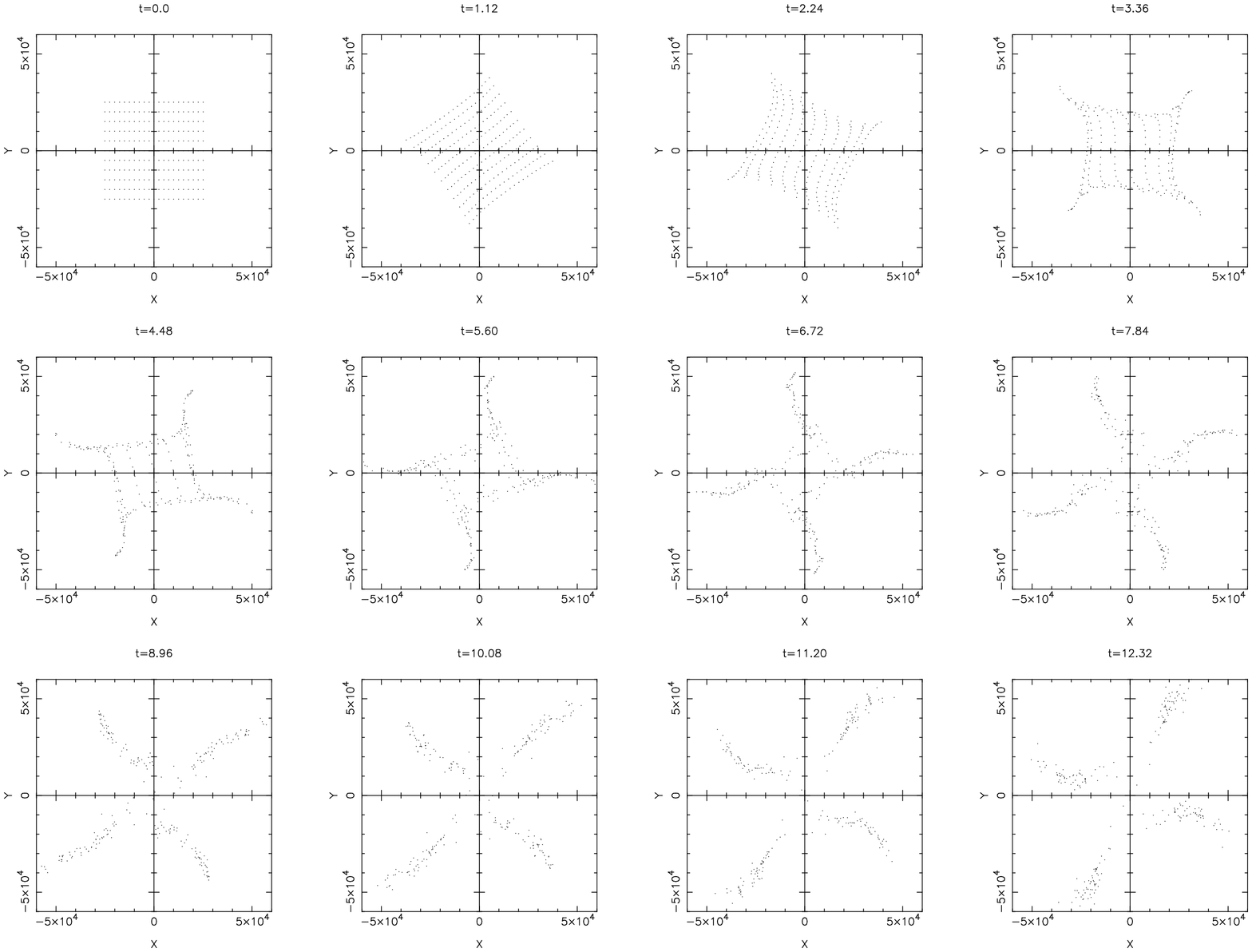,width=8.7cm,height=7.2cm,angle=-90}

\vspace{0.2in}

\epsfig{file=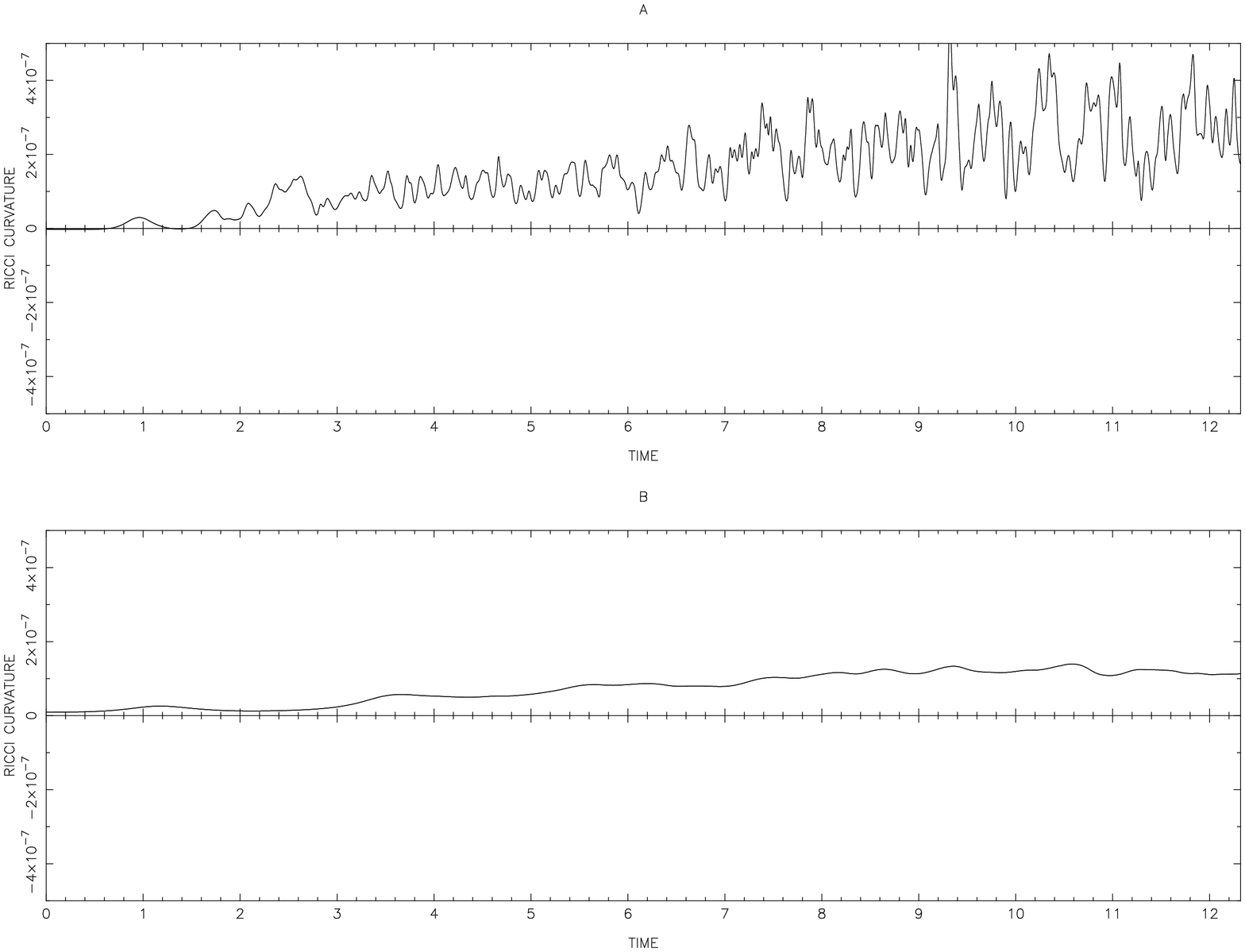,width=8.7cm,height=6.0cm,angle=-90}
\end{flushleft}
\caption{\label{softt}
 The effect of softening:  Same as in  Fig.~\ref{mod7rxy} but when the force law is softened: 
{\bf A} When the softenening radius is 
equal to $10 \%$ the inter-particle distance on the same line,  
{\bf B} same as in {\bf A} but when the softening radius is equal to 
$80 \%$ the  initial inter-particle distance on the same line. The $x-y$ coordinate
evolution (top) correspond to the system in {\bf B} }
\end{figure}

Fig 9A shows the un-averaged Ricci time series
for  a system starting with the same initial conditions as the one in Fig. 8
but when the potential contains a softening parameter (BT formula 2.194) of 
$10 \% $ the initial separation between particles in the same line, while Fig. 9B shows the
the evolution of $r_{\bf u}$ for the same system but when the softening radius 
is about $80 \%$ of the original
separation between particles in the same line.
 First, one notices the
absence of large fluctuations in the value of $r_{\bf u}$. This is to be expected
since  formula (\ref{eq:ru}) no longer contains singularities
at close encounters. The second and  more fundamental effect however is that 
the Ricci curvature is now  positive. This is due to the fact that the last
term on the right hand side is now non-zero and is always positive since it 
corresponds to the density. For small softening (say $1 \%$ of 
inter-particle distance) the original behaviour of $r_{\bf u}$ is recovered,
however as the softening increases this term becomes large. In addition, the
second term in (\ref{eq:ru}), which  in general gives a highly fluctuating 
contribution the average of which is small, starts becoming positive too.  
  
The above means that {\em in the presence of significant softening,
 $N$-body gravitational systems
 no longer  approximate the uniformly hyperbolic Anosov C-systems} because the Ricci curvature is no longer negative on average. That is, there is a fundamental
change of structure in the phase-space 
 of $N$-body systems when softening is introduced. Therefore if one takes the
view that a softened $N$-body system 
is an approximation of the continuum formulation of a problem, one must 
realize that this formulation has solutions which can behave in a different
manner from those of a discrete system. Indeed it was found that the 
evolution of the softened systems differed considerably from the un-softened
ones. For example in Fig. 9 we also show the $x-y$  evolution from the same initial
conditions as in Fig. 8 but when a softening parameter of $80 \%$  the original separation
of particles is present in the force law. Instead of the clustering pattern that was obtained
in the un-softened case, there is now a type of filamentary structure taking its place and the 
system never completely detaches into components. (In fact when the simulation
is continued these filaments are destroyed and one gets a nearly isotropic system.)  Although this system is
still unstable, the dominant mechanism of instability is likely to be different.
It was  observed for example that while in the un-softened system the virial ratio stays close to $1$, in
the softened case it departs significantly from that value, being less than $1$ for the time-scale
of the plot then increasing significantly later. This may suggest that the mechanism of instability here is a parametric instability
related to that effect (Cerruti-Sola \& Pettini 1995). For systems
with intermediate softening a mixture of different mechanisms could
be present. When the negativity of the curvature is dominant, a system
approximates a discrete one. This criteria may be useful in choosing softening
parameters (or laws) in numerical simulations that
 eliminate the effect of close encounters but do not qualitatively change the results.

The difficulty we are encountering here however is a fundamental one.
As long as the system we are considering is discrete and direct collisions
are negligible the offending term in Eq. (\ref{eq:ru}) is zero and the 
Ricci curvature is likely to be negative. As mentioned
in the introduction and in the beginning of section 2.3, this is even more likely to be the case as $N$ increases. On the other hand the smoothing of the
 density distribution leads to positive curvature. The structure of the 
 phase-space (which is the cotangent bundle of the configuration manifold
the curvature of which we are calculating)
in the two cases is then presumably radically different. This suggests that
{\em it is not obvious that large-$N$ but discrete systems are well approximated
by the continuum limit} even if the discreteness noise in the force decreases
as $N$ increases. There appears to be a type of ``phase transition'' in the
structure of the $6N$ phase-space when passing from one limit (large-$N$
but discrete system) to the other (continuous system). This is the same
conclusion we were led to from the general considerations of the introduction.

Obviously these issues will require much more thorough 
investigation which is beyond the scope of this exploratory paper.  
We will however briefly comment on this problem 
from the perspective of the scalar curvature which we now discuss.

\subsection{Scalar curvature}

When the two dimensional curvatures are averaged over all geodesics
that originate at a point of $M$ as well as all possible  directions normal
to the geodesics, one obtains the curvature scalar
\begin{equation}
R=\sum_{\bf u,n} k_{\bf u,n}= \sum_{\bf u} r_{\bf u}.
\end{equation}
With the quantity $R/3N(3N-1)$ replacing $r_{\bf u}/(3N-1)$ in
Equation (\ref{eq:L6}) one can define a corresponding
 instability criterion and an associated time-scale.
This is obviously a more drastic approximation, and because $R$ will not depend
explicitly on the velocities (since we have summed over the directions
in the tangent space $TM$) it cannot contain generic information about
a system and its use for short time characterization of the
dynamics can be especially dangerous (e.g., Cipriani \& Pucacco 1994). Nevertheless, the
similarity of the third term in the formula~(\ref{eq:ru}) 
for the Ricci curvature to
 the expression for $R$ (GS formula (27)) suggests that when
this term is dominant, the scalar curvature may give a reasonable 
description of the dynamics. The fact that estimates of evolutionary
time-scales resulting from the order of magnitude formula of GS agreed within
reasonable bounds with the ones obtained from the computations of
$r_{\bf u}$ suggests that this term may be important in some cases
since that formula is based on an estimate of the scalar curvature.

\begin{figure}
\begin{flushleft}
\epsfig{file=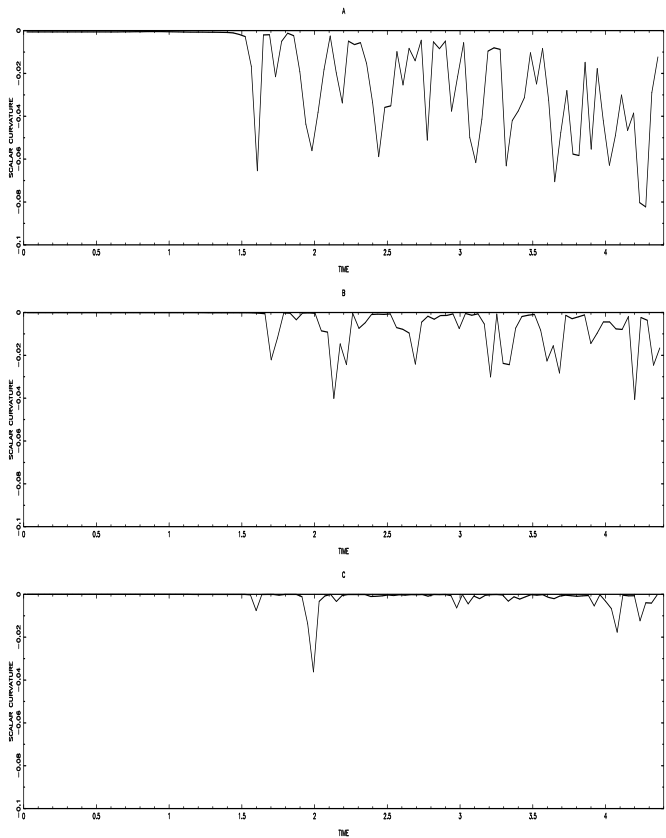,width=8.7cm,height=11.0cm,angle=-90}
\end{flushleft}
\caption{\label{scalar}
 Scalar curvature for the rotating equilibrium models {\bf A} $Scale=1$,
{\bf B} $Scale=10.8$, {\bf C} $Scale=100$ }
\end{figure}

 To
see if this is true,
we have  computed the scalar
curvatures along the motion of the systems described in the previous sections. 
 We have found that using a hundred averaging intervals and filtering
equal to twice the absolute value of the maximum scale in the figures
gives  reasonably smooth time series for $R$ that do not sensitively
depend on the averaging and the filtering threshold used. Fig. 10 shows 
the resultant time series for the  rotating equilibrium models. Clearly, those results agree qualitatively with those 
inferred from the behaviour of the corresponding time series for $r_{\bf
u}$ in Fig. 7
(although differing in some detail). The time-scales also agree (when
$R$ is divided by $3N(3N-1)$ as described above). This agreement
is because out of the three non-zero terms in Eq. (\ref{eq:ru}), the first
is small unless the velocity vectors of particles are almost aligned
with their acceleration vectors and the second is highly fluctuating
and averages to very small values (this is because it describes
the variations of the force at the particle positions which, as one would expect, 
 depends sensitively on the position of close neighbours). This leaves the third term which is the one that appears in the formula for the scalar
curvature.

The scalar curvature --- like the Ricci curvature ---
 also takes positive values when significant softening is used. As was noted in the introduction,
 the continuum approximation with a steady state potential
reduces the $N$-body
problem to $N$ one-particle problems in a given potential. In this context,
the above discrepancy between the value of the curvatures of 
softened and un-softened cases
 may perhaps be understood by noticing that the
scalar curvature  for $N=1$  is {\em always} positive  no matter what the
potential is (assuming positive density). The 6-dimensional phase-space
is therefore far from approximating a hyperbolic structure like that of C-systems.
Even in this case, chaos may still occur (some two dimensional curvatures
may still be negative and there is also the possibility of parametric
instability: ARN; P93) but it will not likely be as robust or widespread 
as in systems with negative curvature.

\section{Conclusions}

The assumptions of classical non-evolutionary
galactic dynamics may not be satisfied 
in the presence of significant amount of chaos (section 1). This makes
it important to develop and to test methods characterizing such chaotic
behaviour. Geometric  methods have the double advantage of being 
{\em local} and comparing {\em normal} deviations between trajectories of dynamical
systems and not the total divergence of temporal states.
 For higher
dimensional systems these properties may be important in distinguishing
between true phase space mixing which can be accompanied by changes
in the physical characteristics of a system and phase mixing which conserves the action 
variables.
Geometric methods
are  therefore better suited for studies of the short time evolution of
galaxies than other (more traditional) measures of chaos (section 2 and 3.1). 

In  large dimensional systems it may not be 
practical to determine the stability of a trajectory to all possible perturbations.
The next best thing is to determine the average 
divergence of trajectories due to random perturbations.
One way of doing this is by examining the  Ricci curvature of the Lagrangian 
configuration manifold of a dynamical system (section 2.3). 
To check the effectiveness of such an approach for $N$-body gravitational systems
we have calculated the Ricci curvature for several small
$N$-body  systems integrated with high precision (section 3.2 and 4).
 The results of these experiments show that:
\begin{enumerate}
\item
When properly averaged to get rid of the contributions of close encounters the Ricci
curvature is almost always negative, confirming that gravitational systems
are unstable (e.g., Miller 1964; Goodman et al. 1993; Kandrup et al. 1994)
and that the main mechanism of instability is the negativity
of the curvature of the configuration manifold as predicted by
Gurzadyan \& Savvidy (1984,1986) and Kandrup (1990a,1990b).
\item
The Ricci curvature is more negative (hence predicts shorter evolutionary time-scales)
 when a system develops pronounced  macroscopic instabilities (e.g., plasma
type collective instabilities). In this case 
the rates of spatial macroscopic evolution of the different systems both relative 
to each other and in terms of evolution time-scales
was reasonably well described
by the time-scales derived on the basis of the Ricci curvature calculations.
\item
When expressed in terms of  dynamical times,
evolutionary time-scales appeared to be slightly longer for systems starting from virial
equilibrium than those starting from a virial ratio less than one. However, since
when the kinetic energy varies significantly, chaotic behaviour cannot be described
by the negativity of the Ricci curvature alone (Cerruti-Sola \& Pettini 1995) this
conclusion remains to be confirmed. (However  in the case of 
large $N$-body  systems near
virial equilibrium  the objections to the use of the Ricci curvature as outlined in the  
 aforementioned paper are not likely to be important since the second and third term of 
their Eq. (26) are then very small. By using the Ricci curvature 
and eliminating large fluctuations due to close encounters  we have therefore implicitly assumed that it is the  instability in that 
(large-$N$) limit that interests us and not effects due to small 
scale fluctuations.)
\item
In the  presence of significant (but not very large) softening, the Ricci curvature 
becomes positive. This probably means that the phase-space structure of softened
systems is radically different from that of point particles. This has consequences for 
the interpretation of results of numerical simulations. More fundamentally however
this result may be interpreted to mean that {\em there is no continuous transition from large-$N$ discrete $N$-body  systems to continuous ones}.
 It may therefore explain  why large $N$-body  spherical systems
are found to approximate exponentially unstable C-systems while it is known that motion
in smooth spherical potentials is separable. More work however is needed to fully understand
the meaning of this effect and its possible consequences.
\item
Results derived on the basis of the scalar curvature agreed, in general,
with those obtained from the evolution of the Ricci curvature. This shows
the instability to be quite a robust phenomenon.

\end{enumerate}

We have not looked at properties such as energy relaxation here. 
As was mentioned in the introduction  this can have different
time-scales from that of the instability of trajectories. 
To see how this may be the case we consider the change in the
Hamiltonian $H=T+V$ of a test particle $i$ in an $N$-body system. Using
the Hamiltonian equations we find that along the motion this is given by
\begin{equation}
\frac{d H_{i}}{d t}=\frac{\partial H_{i}}{\partial t}=\frac{\partial
V_{i}}{\partial t}
\end{equation}
with the interaction potential $V_{i}$ given by
\begin{equation}
V_{i}=\sum_{j} V_{ij}
\end{equation}
being a function of the positions of the remaining $N-1$ particles which are of
course time dependent functions of the initial conditions. The change in
energy of particle $i$ along its path is then given by
\begin{equation}
\int \frac{\partial V_{i}}{\partial t} dt 
= \int \sum_{j} \frac{\partial V_{ij}}{\partial t} dt
= \sum_{j} \int \frac{\partial V_{ij}}{\partial t} dt,
\end{equation}
which is the sum of the energy changes due to the individual
interactions and therefore could proceed on time-scales similar
to that given in (1). 
This does not of course mean that energy relaxation cannot be  enhanced
 for  systems consisting of particles 
 with different masses or those out of virial
equilibrium or where collective motion  or large scale inhomogeneity
or anisotropy 
occurs. In these cases chaotic
behaviour can be important for energy relaxation. Indeed there is some evidence that in
some situations two body relaxation estimates are inaccurate even for energy
relaxation (El-Zant 1996a). In general however there may be phase-space
``barriers''  across which diffusion is slow. This may
 prevent some quantities from relaxing even when chaos
is present (discussions of the issue of phase-space transport in
higher dimensional Hamiltonian systems can be found in Wiggins 1991 or Benettin 1994). 
Nevertheless, chaos implies exponential divergence {\em normal}
to the phase-space trajectory leading to diffusion in at least some of the action variables which
determine the physical characteristics of a system (as opposed to phase mixing 
which conserves the action variables).
It therefore has important consequences for the behaviour
of dynamical systems as was stressed in the introductory section of this paper. 
This has long been generally recognised in various branches of Physics and Mathematics (e.g., Sagdeev et al. 1988)
but not always in stellar dynamics. The question therefore is 
{\em how widespread is the chaotic behaviour in  realistic $N$-body realizations
of the different galaxy types and
what are the exponentiation time-scales}. This is what we hope to find out using the
method examined in this paper.

\subsection*{Acknowledgments}
 I would like to thank Prof. R.J. Tayler for constructive comments on 
an earlier version of this paper, Prof. V.G. Gurzadyan  for enlightening
 discussions and some important suggestions, 
 Simon Goodwin for  helping me with the English and the 
referee Daniel Pfenniger for comments that helped improve the presentation.
I would also like to thank Henry Kandrup for a useful discussion that helped
clarify some of my ideas on the subject of this paper and Douglas Heggie for 
helpful communication. 
This work was undertaken while the author
 was supported by a Foreign and Commonwealth Office Chevening scholarship.

\end{document}